\documentstyle[12pt]{article}
\advance\topmargin by -1.6cm

\begin{document}

\newpage

\newenvironment{eqn}
{\begin{equation}\begin{array}}{\end{array}\end{equation}{}} 
\newenvironment{eq}
{\[\begin{array}}{\end{array}\]{}}


\let\rvec=\vec        
\let\Sup=\sup        
\let\Top=\top        
\let\pfi=\phi

 
\def\sm{\setminus}   \def\({\Bigl(}
\def\){\Bigr)}    \def\|{\Big|}
\def\then{~\Rightarrow~}   \def\o{\circ}
\def\m{\bullet}    \def\x{\times}
\def\oo{\infty}   \def\ox{\otimes}
\def\Ox{\bigotimes} \def\OX{\displaystyle\bigotimes}
\def\pl{{~\oplus~}}
\def\Pl{\bigoplus}
\def\PL{\displaystyle \bigoplus}
\def\SUM{\displaystyle \sum}
\def\PROD{\displaystyle \prod}
\def\BIGUPLUS{\displaystyle \biguplus}
\def\mid{\big\bracevert}
\def\Cup{\bigcup}
\def\Cap{\bigcap}
\def\CUP{\displaystyle \Cup}
\def\CAP{\displaystyle \Cap}
\def\sub{\subseteq}
\def\subnoteq{\subset}
\def\sup{\supseteq}
\def\supnoteq{\supset}
\def\and{\wedge}
\def\And{\bigwedge}
\def\AND{\displaystyle\bigwedge}
\def\od{\vee}
\def\Od{\bigvee}
\def\OD{\displaystyle\bigvee}
\def\tria{\triangleleft}
\def\dia{\diamond}
\def\vel{\dot +}
\def\rin{{\,\in\kern-.42em\in}}
 \def\der{{\,{\rm der}\,}}
 \def\diag{{\,{\rm diag}\,}}
\def\sign{{\,{\rm sign}\,}}
\def\rang{\,{\rm rang}\,}
\def\rank{\,{\rm rank}}
\def\spec{\,{\rm spec}\,}
\def\rep{\,{\rm rep}\,}
\def\supp{\,{\rm supp }\,}
\def\tr{{\,{\rm tr }\,}}
\def\det{\,{\rm det }\,}
\def\id{\,{\rm id}}
\def\is{\,{\rm is}}
\def\as{\,{\rm as}}
\def\op{{\rm op}}
\def\co{{\rm co}}
\def\Re{\,{\rm Re}}
\def\Im{\,{\rm Im}}
\def\Inv{\,\hbox{INV}}
\def\INV{\,\hbox{INV}}
\def\CLIFF{\,\hbox{CLIFF}}
\def\CLAG{\,\hbox{CLAG}}
\def\FIX{\,\hbox{FIX}}
\def\STAB{\,\hbox{STAB}}
\def\Int{\,{\rm Int}\,}
\def\Ad{\,{\rm Int}\,}
\def\ad{{\,{\rm ad}\,}}
\def\centr{\,{\rm centr}\,}
\def\Kern{\,{\rm kern}\,}
\def\Bild{\,{\rm Bild}\,}
\def\Imag{\,{\rm imag}\,}
\def\unit{\,{\rm unit}\,}
\def\caus{{\scriptsize{\rm caus}}}
\def\deg{\,{\rm deg}\,}
\def\db{{\scriptsize{\rm doub}}}
\def\sqsub{\sqsubseteq}
\def\sqsup{\sqsupseteq}
\def\voll{\spadesuit}
\def\leer{\heartsuit}
\def\card{\ro{\,card\,}}
\def\X{\hbox{\Large$\times$}}
\def\sx{~\rvec\x~\!}
\def\weights{{{\bf weights}\,}}
\def\irrep{{{\bf irrep\,}}}
\def\rep{{{\bf rep\,}}}
\def\ndecrep{{{\bf ndecrep\,}}}
\def\meas{{\bf meas}\,}

\def\A{{\,{\rm A\kern-.55emA}}}
\def\B{{\,{\rm I\kern-.2emB}}}
\def\C{{\,{\rm I\kern-.55emC}}}
\def\E{{\,{\rm I\kern-.2emE}}}
\def\G{{\,{\rm I\kern-.55emG}}}
\def\H{{{\rm I\kern-.2emH}}}
\def\I{{\,{\rm I\kern-.2emI}}}
\def\K{{\,{\rm I\kern-.2emK}}}
\def\L{{\,{\rm I\kern-.2emL}}}
\def\M{{\,{\rm I\kern-.16emM}}}
\def\N{{\,{\rm I\kern-.16emN}}}
\def\Q{{\,{\rm I\kern-.5emQ}}}
\def\R{{{\rm I\kern-.2emR}}}
\def\S{{\,{\rm I\kern-.42emS}}}
\def\T{{\,{\rm I\kern-.37emT}}}
\def\UU{{\,{\rm I\kern-.51emU}}}
\def\Z{{\,{\rm Z\kern-.32emZ}}}

\def\p{\partial}
\def\cd{\nabla}
\def\pd#1#2{{\partial#1\over{\partial #2}}}
 
 
\def\al{\alpha}  \def\be{\beta} \def\ga{\gamma}
\def\de{\delta}  \def\ep{\epsilon}  \def\ze{\zeta}
\def\th{\theta}   \def\vth{\vartheta} \def\io{\iota}
\def\ka{\kappa}   \def\la{\lambda}   \def\si{\sigma}
\def\De{\Delta}   \def\om{\omega} \def\Om{\Omega}
\def\phi{\varphi} 
 \def\Ga{\Gamma}  \def\Th{\Theta}
\def\Si{\Sigma}    \def\La{\Lambda}

 
\def\kat{\underline{\bf kat}}
\def\set{\underline{\bf set}}
\def\oset{\underline{\bf oset}}
\def\latt{\underline{\bf latt}}
\def\struc#1{\underline{\bf struc}_{#1}}
\def\mon{\underline{\bf mon}}
\def\grp{\underline{\bf grp}}
\def\abgrp{\underline{\bf abgrp}}
\def\rng{\underline{\bf rng}}
\def\mod#1{\underline{\bf mod}_{#1}}
\def\modf#1{\underline{\bf mod}_{#1}^{{\bf frei}}}
\def\vec#1{\underline{\bf vec}_{#1}}
\def\tvec#1{\underline{\bf tvec}_{#1}}
\def\nvec#1{\underline{\bf nvec}_{#1}}
\def\cnvec#1{\underline{\bf cnvec}_{#1}}
\def\svec#1{\underline{\bf svec}_{#1}}
\def\csvec#1{\underline{\bf csvec}_{#1}}
\def\ag#1{\underline{\bf ag}_{#1}}
\def\aag#1{\underline{\bf aag}_{#1}}
\def\naag#1{\underline{\bf naag}_{#1}}
\def\cnaag#1{\underline{\bf cnaag}_{#1}}
\def\lag#1{\underline{\bf lag}_{#1}}
\def\lrg{\underline{\bf lrg}}
\def\top{\underline{\bf top}}
\def\utop{\underline{\bf utop}}
\def\dif#1{\underline{\bf dif}_{#1}}
\def\lgrp#1{\underline{\bf lgrp}_{#1}}
\def\bdl#1{\underline{\bf bdl}_{#1}}
\def\mes{\underline{\bf mes}}
\def\mas{\underline{\bf mas}}
 
 
\def\katg#1{{\bf kat}(#1)}
\def\setg#1{{\bf set}(#1)}
\def\mong#1{{\bf mon}(#1)}
\def\grpg#1{{\bf grp}(#1)}
\def\abgrpg#1{{\bf abgrp}(#1)}
\def\rngg#1{{\bf rng}(#1)}
\def\modg#1#2{{\bf mod}_{#1}(#2)}
\def\vecg#1#2{{\bf vec}_{#1}(#2)}
\def\agg#1#2{{\bf ag}_{#1}(#2)}
\def\aagg#1#2{{\bf aag}_{#1}(#2)}
\def\lagg#1#2{{\bf lag}_{#1}(#2)}
\def\topg#1{{\bf top}(#1)}
\def\difg#1#2{{\bf dif}_{#1}(#2)}
\def\mesg#1{{\bf mes}(#1)}
 
 
\def\katm#1#2{{\bf kat}(#1,#2)}
\def\setm#1#2{{\bf set}(#1,#2)}
\def\setom#1#2#3{{\bf set}_{#1}(#2,#3)}
\def\osetm#1#2{{\bf oset}(#1,#2)}
\def\lattm#1#2{{\bf latt}(#1,#2)}
\def\strucm#1#2#3{{\bf struc}_{#1}(#2,#3)}
\def\monm#1#2{{\bf mon}(#1,#2)}
\def\grpm#1#2{{\bf grp}(#1,#2)}
\def\abgrpm#1#2{{\bf abgrp}(#1,#2)}
\def\rngm#1#2{{\bf rng}(#1,#2)}
\def\modm#1#2#3{{\bf mod}_{#1}(#2,#3)}
\def\vecm#1#2#3{{\bf vec}_{#1}(#2,#3)}
\def\tvecm#1#2#3{{\bf tvec}_{#1}(#2,#3)}
\def\nvecm#1#2#3{{\bf nvec}_{#1}(#2,#3)}
\def\cnvecm#1#2#3{{\bf cnvec}_{#1}(#2,#3)}
\def\svecm#1#2#3{{\bf svec}_{#1}(#2,#3)}
\def\csvecm#1#2#3{{\bf csvec}_{#1}(#2,#3)}
\def\agm#1#2#3{{\bf ag}_{#1}(#2,#3)}
\def\aagm#1#2#3{{\bf aag}_{#1}(#2,#3)}
\def\naagm#1#2#3{{\bf naag}_{#1}(#2,#3)}
\def\cnaagm#1#2#3{{\bf cnaag}_{#1}(#2,#3)}
\def\lagm#1#2#3{{\bf lag}_{#1}(#2,#3)}
\def\topm#1#2{{\bf top}(#1,#2)}
\def\utopm#1#2{{\bf utop}(#1,#2)}
\def\difm#1#2#3{{\bf dif}_{#1}(#2,#3)}
\def\lgrpm#1#2#3{{\bf lgrp}_{#1}(#2,#3)}
\def\bdlm#1#2#3{{\bf bdl}_{#1}(#2,#3)}
\def\mesm#1#2{{\bf mes}(#1,#2)}
\def\masm#1#2{{\bf mas}(#1,#2)}
 
 
\def\kati#1#2{\stackrel{\rm o}{\bf kat}(#1,#2)}
\def\seti#1#2{\stackrel{\rm o}{\bf set}(#1,#2)}
\def\moni#1#2{\stackrel{\rm o}{\bf mon}(#1,#2)}
\def\grpi#1#2{\stackrel{\rm o}{\bf grp}(#1,#2)}
\def\abgrpi#1#2{\stackrel{\rm o}{\bf abgrp}(#1,#2)}
\def\rngi#1#2{\stackrel{\rm o}{\bf rng}(#1,#2)}
\def\modi#1#2#3{\stackrel{\rm o}{\bf mod}_{#1}(#2,#3)}
\def\veci#1#2#3{\stackrel{\rm o}{\bf vec}_{#1}(#2,#3)}
\def\tveci#1#2#3{\stackrel{\rm o}{\bf tvec}_{#1}(#2,#3)}
\def\nveci#1#2#3{\stackrel{\rm o}{\bf nvec}_{#1}(#2,#3)}
\def\cnveci#1#2#3{\stackrel{\rm o}{\bf cnvec}_{#1}(#2,#3)}
\def\sveci#1#2#3{\stackrel{\rm o}{\bf csvec}_{#1}(#2,#3)}
\def\csveci#1#2#3{\stackrel{\rm o}{\bf csvec}_{#1}(#2,#3)}
\def\agi#1#2#3{\stackrel{\rm o}{\bf ag}_{#1}(#2,#3)}
\def\aagi#1#2#3{\stackrel{\rm o}{\bf aag}_{#1}(#2,#3)}
\def\naagi#1#2#3{\stackrel{\rm o}{\bf naag}_{#1}(#2,#3)}
\def\cnaagi#1#2#3{\stackrel{\rm o}{\bf cnaag}_{#1}(#2,#3)}
\def\lagi#1#2#3{\stackrel{\rm o}{\bf lag}_{#1}(#2,#3)}
\def\topi#1#2{\stackrel{\rm o}{\bf top}(#1,#2)}
\def\utopi#1#2{\stackrel{\rm o}{\bf utop}(#1,#2)}
\def\difi#1#2#3{\stackrel{\rm o}{\bf dif}_{#1}(#2,#3)}
\def\lgrpi#1#2#3{\stackrel{\rm o}{\bf lgrp}_{#1}(#2,#3)}
\def\bdli#1#2#3{\stackrel{\rm o}{\bf bdl}_{#1}(#2,#3)}
\def\mesi#1#2{\stackrel{\rm o}{\bf mes}(#1,#2)}
\def\masi#1#2{\stackrel{\rm o}{\bf mas}(#1,#2)}
 

\def\GL{{\bf GL}}  
\def\SL{{\bf SL}}
\def\U{{\bf U}} 
\def \UL{{\bf UL}} 
\def\O{{\bf O}}   
\def\SU{{\bf SU}} 
\def\SD{{\bf SD}} 
\def\SO{{\bf SO}}
 \def\Sp{{\bf Sp}} 
 \def\USp{{\bf USp}}
 \def\D{{\bl D}}
\def\AL{{\bf AL}}

 
\def\norm#1{\parallel #1\parallel}
\def\Ffo#1{\angle{#1}_{{}_{{\rm F}}}   }  
\def\Fnorm#1{\norm {#1}_{{}_{{\rm F}}}   }  
\def\TFfo#1{\angle{#1}_{\bl T}   }  
\def\Hfo#1{\angle{#1}_{{}_{{\rm H}}}   }  
\def\d#1{{\check{#1}}}
\def\angle#1{\langle#1\rangle}
\def\Angle#1{\Bigl\langle#1\Bigr\rangle}
\def\aangle#1{\langle\!\langle#1\rangle\!\rangle}
\def\slash#1{#1\!\!\!\!/}
\def\qform#1{\prec#1\succ}
\def\lrvec#1{
{\textstyle^{^{^{\leftrightarrow}}}\displaystyle\!\!\!\!\!#1}}
\def\rstate#1{|#1\rangle}
\def\lstate#1{\langle#1|}
\def\brack#1{\lbrack#1\rbrack}
\def\Brack#1{\Bigl\lbrack#1\Bigr\rbrack}
\def\mini#1{{\scriptstyle#1\displaystyle}}
\def\ty#1{{\tt #1}}
\def\ro#1{{\rm #1}}
\def\bl#1{{\bf {#1}}}
\def\cl#1{{\cal #1}}
\def\ul#1{\underline{#1}}
\def\uul#1{\ul{\ul{#1}}}
\def\ol#1{\overline{#1}}
\def\brace#1{\lbrace#1\rbrace}
\def\Brace#1{\Bigl\lbrace#1\Bigr\rbrace}

 
\def\dprod#1#2{\langle#1,#2\rangle}
\def\sprod#1#2{\langle#1|#2\rangle}
\def\dyn#1#2{\matrix{#1\cr \o \cr \om_{#2}\cr}}
\def\com#1#2{\lbrack#1,#2\rbrack}
\def\Com#1#2{ \Bigl\lbrack #1,#2 \Bigr\rbrack }
\def\acom#1#2{\{#1,#2\}}
\def\Acom#1#2{\Bigl\{#1,#2\Bigr\}}
\def\bra#1#2{\lbrack\!\lbrack#1,#2\rbrack\!\rbrack}
\def\Bra#1#2{
 \Bigl\lbrack\>\!\!\!\Bigl\lbrack#1,#2\Bigr\rbrack\>\!\!\!\Bigr\rbrack}
 
 
\def\expv#1#2#3{\langle#1|#2|#3\rangle}
 
 
\def\map{\longrightarrow}
\def\inmap{\hookrightarrow}
\def\lrmap{\leftrightarrow}
\def\llrmap{\lmap\hskip-4mm\map}
\def\lmap{\longleftarrow}
\def\dmap{\Big\downarrow}
\def\umap{\Big\uparrow}
\def\fun{~\rule[.75mm]{9mm}{.5mm}\!\!\!\!\succ} 
\def\wigglyfun{\sim\!\sim\!\sim\!\sim\!\sim\!\succ}
\def\mape{\longmapsto}
\def\lmape{\longleftarrow\!\!\!{\tiny\vrule}}
\def\dmape{\vcenter{\hrule\hbox{$\Big\downarrow$}}}
\def\umape{\vcenter{\hbox{$\Big\uparrow$}\hrule}}
\def\fune{~\rule[-.1mm]{.5mm}{2mm}\rule[.75mm]{9mm}{.5mm}\!\!\!\!\succ}    
\def\wigglyfune{\vert\!\!\!\fun}
\def\downmapl#1#2#3{\matrix{&#1\cr{\scriptstyle #2}&\dmap\cr&#3\cr}}
\def\downmapr#1#2#3{\matrix{#1&\cr\dmap&{\scriptstyle#2}\cr#3&\cr}}
\def\upmap#1#2#3{\matrix{#3&\cr\umap&{\scriptstyle #2}\cr#1&\cr}}
\def\downmape#1#2{\matrix{#1\cr\dmape\cr#2\cr}}
\def\upmape#1#2{\matrix{#2\cr\umape\cr#1\cr}}
\def\diagr#1#2#3#4#5#6{\matrix{  \noalign{\vskip5mm}
&&{\scriptstyle #4}& \cr
                                   &#1   &\map                     &#2\cr
                   {\scriptstyle #6}&\dmap&\nearrow                 &  \cr
                                   &#3   &~~~~^{\scriptstyle#5}  &  \cr
  \noalign{\vskip5mm}                }}
 
\def\diagre#1#2#3{\matrix{\noalign{\vskip5mm}
                                      &&\cr
                               #1&\mape&#2\cr
                           \dmape&\nearrow&\cr
                             #3&          &\cr
 \noalign{\vskip5mm}            }}
 
\def\Diagr#1#2#3#4#5#6#7#8{\matrix{\noalign{\vskip5mm}
      &              &{\scriptstyle #5}&              &     \cr
      & #1           & \map           & #2           &     \cr
{\scriptstyle #8}   &\dmap         &    &\dmap  &{\scriptstyle#6} \cr
      & #4           & \map           & #3           &     \cr
      &              &{\scriptstyle#7}&              &     \cr
\noalign{\vskip5mm}             }}

\def\dualDiagr#1#2#3#4#5#6#7#8{\matrix{\noalign{\vskip5mm}
      &              &{\scriptstyle #5}&              &     \cr
      & #1           & \lmap           & #2           &     \cr
{\scriptstyle #8}   &\umap         &    &\umap  &{\scriptstyle#6} \cr
      & #4           & \lmap           & #3           &     \cr
      &              &{\scriptstyle#7}&              &     \cr
\noalign{\vskip5mm}             }}

\def\Diagrlrud#1#2#3#4#5#6#7#8{\matrix{\noalign{\vskip5mm}
      &              &{\scriptstyle #5}&              &     \cr
      & #1           & \llrmap           & #2           &     \cr
{\scriptstyle #8}   &\updownarrow& &\updownarrow&{\scriptstyle#6} \cr
      & #4           & \llrmap           & #3           &     \cr
      &              &{\scriptstyle#7}&              &     \cr
\noalign{\vskip5mm} 
            }}
\def\Diagrlr#1#2#3#4#5#6#7#8{\matrix{\noalign{\vskip5mm}
      &              &{\scriptstyle #5}&              &     \cr
      & #1           & \lrmap           & #2           &     \cr
{\scriptstyle #8}   &\updownarrow& &\updownarrow&{\scriptstyle#6} \cr
      & #4           & \lrmap           & #3           &     \cr
      &              &{\scriptstyle#7}&              &     \cr
\noalign{\vskip5mm}             }}

\def\Diagrud#1#2#3#4#5#6#7#8{\matrix{\noalign{\vskip5mm}
      &              &{\scriptstyle #5}&              &     \cr
      & #1           & \map           & #2           &     \cr
{\scriptstyle #8}   &\updownarrow& &\updownarrow&{\scriptstyle#6} \cr
      & #4           & \map           & #3           &     \cr
      &              &{\scriptstyle#7}&              &     \cr
\noalign{\vskip5mm}             }}

\def\dualDiagr#1#2#3#4#5#6#7#8{\matrix{\noalign{\vskip5mm}
      &              &{\scriptstyle #5}&              &     \cr
      & #1           & \lmap           & #2           &     \cr
{\scriptstyle #8}   &\umap         &    &\umap  &{\scriptstyle#6} \cr
      & #4           & \lmap           & #3           &     \cr
      &              &{\scriptstyle#7}&              &     \cr
\noalign{\vskip5mm}             }}

\def\Diagrup#1#2#3#4#5#6#7#8{\matrix{\noalign{\vskip5mm}
      &              &{\scriptstyle #5}&              &     \cr
      & #1           & \map           & #2           &     \cr
{\scriptstyle #8}   &\umap         &    &\umap  &{\scriptstyle#6} \cr
      & #4           & \map           & #3           &     \cr
      &              &{\scriptstyle#7}&              &     \cr
\noalign{\vskip5mm}             }}

\def\Diagrdownup#1#2#3#4#5#6#7#8{\matrix{\noalign{\vskip5mm}
      &              &{\scriptstyle #5}&              &     \cr
      & #1           & \map           & #2           &     \cr
{\scriptstyle #8}   &\dmap         &    &\umap  &{\scriptstyle#6} \cr
      & #4           & \map           & #3           &     \cr
      &              &{\scriptstyle#7}&              &     \cr
\noalign{\vskip5mm}             }}

\def\Diagre#1#2#3#4{\matrix{ \noalign{\vskip0mm}
                                             &&\cr
         #1      &\mape       &   #2            \cr
     \dmape      &            &  \dmape         \cr
         #4      &\mape       &   #3            \cr
\noalign{\vskip5mm}             }}
 
\def\Triagr#1#2#3#4#5#6{
#1
\matrix{{\scriptstyle#4}\cr\swarrow\cr\nwarrow\cr{\scriptstyle#6}\cr}
\matrix{#2&  \cr\dmap&{\scriptstyle#5}\cr#3& \cr}
}
 
\def\abb#1#2#3{{\scriptsize\matrix{
  &#3  &  \cr
#1&\map&#2\cr
&~&\cr}}}
 
\def\erw#1#2#3#4#5{{\scriptsize\matrix{
  &#4  &  & #5 &\cr
#1&\map&#2&\map&#3\cr
  &~  &  &  &\cr
}}}


\begin{titlepage} 
\hfill MPI-PhT/00-41 


\vskip25mm
\centerline{\bf RESIDUAL REPRESENTATIONS}
 
\centerline{\bf  OF SPACETIME}
\vskip1cm
\centerline{
Heinrich Saller\footnote{\scriptsize 
hns@mppmu.mpg.de}  
}
\centerline{Max-Planck-Institut f\"ur Physik and Astrophysik}
\centerline{Werner-Heisenberg-Institut f\"ur Physik}
\centerline{M\"unchen}
\vskip25mm

\centerline{\bf Abstract}

\vskip5mm

Spacetime is modelled by binary relations - 
by the classes of the automorphisms $\GL(\C^2)$ of a complex
2-dimensional vector space with respect to the definite unitary    
subgroup $\U(2)$. In extension of Feynman propagators for particle quantum
fields representing only the tangent spacetime structure, global spacetime
representations are given, formulated as residues using 
energy-momentum distributions with the invariants as singularities. The associated 
quantum fields are characterized by
two invariant masses - for time and position - supplementing  the one mass for the
definite unitary 
particle sector with another mass for the indefinite unitary interaction sector
without asymptotic particle interpretation.


\end{titlepage}

{\small \tableofcontents}

\newpage

\section{Introduction}

Quantum theory starts with operations\cite{FINK}.
 An experiment for quantum structures  probes
a `diagonalization' of the operator under question,
e.g. of a time and position translation or of a rotation or of  a 
charge transformation, 
with the eigenvalues as possible experimental 
results, e.g. energy and momenta and mass or spin or a charge number resp.
Therewith, 
I shall take the radical point of view that all relevant 
mathematical structures and tools
used in quantum theories have to have an interpretation in terms of operations, of 
monoids, groups and  algebras, especially of real Lie groups and Lie algebras,
realized and represented as acting upon sets, especially upon complex
vector spaces with a reality  defining conjugation.
Representation theory gives the  irreducible 
and  - for linear structures - also the nondecomposable action spaces.
Almost all functions, relevant for physics, can be interpreted as arising from
representation structures\cite{VIL}.

Physical events
represent spacetime operations,
e.g. translations, rotations and boosts.
A quantum mechanical dynamics,  implemented by $iH$ (Hamiltonian
$H$ with eigenvalues $E\in\R$) as basis for  the time translation Lie algebra $\R$,
is a re\-pre\-sen\-ta\-tion of the causal time
group $\D(1)=\exp\R$, irreducible for $e^{ti E }\in\U(1)$, 
e.g. for the harmonic
oscillator or for creation and annihilation operators in quantum particle
fields. In the Schr\"odinger picture 
the  time representations in $\U(1)$ are realized on a
Hilbert space with the scalar product (probability amplitudes) 
induced by the time representing $\U(1)$.
The wave functions come as  position translation 
representation matrix elements,  e.g. the scattering and bound state
 wave functions $\psi(r)$ in 
rotation symmetric problems with 
$r\psi(r)\sim e^{\pm ir|Q|},~e^{-r|Q|}$
as compact $\U(1)$ and noncompact $\D(1)$-representations resp. of 
the radial translation monoid $\R^+$. In quantum mechanics
the time translation eigenvalue   $iE$ (energy $E$) and the 
 position translation eigenvalue $Q$
are in a unique correpondence: E.g., for a
constant potential $V_0$ with $-{Q^2\over2}=E-V_0$ the scattering
case is given by $E>V_0$ with imaginary eigenvalues $\pm i|Q|$ 
and momentum $|Q|$ whereas the
bound states come with $E<V_0$ - there $|Q|$ cannot be interpreted as 
momentum.

In analogy to the dynamics for time $\D(1)=\exp\R$  
the re\-pre\-sen\-ta\-tions\footnote{\scriptsize
$\irrep G$ and $\rep G$ denotes the (irreducible) representation classes of a
group $G$.} of the globally symmetric 
manifold $\D(2)=\exp\R^4$  
as spacetime model\cite{S97,S991}  (discussed in more detail below) 
with the Minkowski  translations as tangent space
 $\R^4$ 
will be considered as possible candidates for a spacetime 
dynamics
\begin{eq}{rll}
\hbox{time  dynamics}:&\rep \D(1)&\hbox{with }\D(1)=\GL(\C )/\U(1)\cr
\hbox{spacetime dynamics}:&\rep \D(2)
&\hbox{with }\D(2)\cong\GL(\C^2 )/\U(2)\cr
\end{eq}The spacetime manifold $\D(2)=\D(\bl1_2)\x\SD(2)$ contains 
as factor for the causal group $\D(1)$ the
rank 1 position   manifold $\SD(2)\cong\SO_0(1,3)/\SO(3)$
with another Cartan subgroup $\SO_0(1,1)\cong\exp \R$.
An independent realization of both 
factors in the Cartan subgroups 
$\D(1)\x\SO_0(1,1)$ of the rank 2 
spacetime manifold $\D(2)$ is characterized by
two continuous invariants.

For particles with mass $m$, the energy-momenta $(q_0,\rvec q)$ 
as eigenvalues for
spacetime translations $(x_0,\rvec x)$ 
are on shell, i.e.  $q^2=m^2$. With Wigner\cite{WIG}, 
particle quantum fields implement definite unitarily the  Poincar\'e Lie algebra 
with the  mass $m^2=q_0^2-\rvec q^2$ as the 
translation eigenvalue.
In the following the off shell structures
of a propagator, i.e. for $q^2\ne m^2$,
 will be extended for a complete realization
of rank 2 spacetime $\D(2)$ with its two noncompact invariants.

Representation matrix elements\footnote{\scriptsize 
In the following the short `re\-pre\-sen\-ta\-tion' can stand for 
the more correct 're\-pre\-sen\-ta\-tion matrix
element(s)'.} of a real  Lie group 
are analytic functions on this group 
\begin{eq}{l}
D:G\map \C,~~g( x)\mape D( x)
\end{eq}e.g. ${\rvec x\over  r }i\sin  r $
for compact spin $\SU(2)$
or $\cos xm$ for compact axial rotations $\U(1)$ or 
both $\cos xm$ and $t\cosh xm$  for noncompact
time $\D(1)$.
According to the Peter-Weyl theorem\cite{PWEYL,FOLL}, the span
of the irreducible re\-pre\-sen\-ta\-tion matrix elements  of a compact Lie group
is dense in the continuous  functions
on this group.

In a harmonic analysis, 
re\-pre\-sen\-ta\-tion matrix elements of a group can be written as  
Fourier transforms of  distributions of their  Lie
algebra forms, e.g. of energies or angular momenta values,
where the re\-pre\-sen\-ta\-tion characterizing invariants come as singularities,
i.e. as poles of the distributions.
This defines the concept of residual re\-pre\-sen\-ta\-tions.
In the following,  familiar algebraic re\-pre\-sen\-ta\-tion
concepts\cite{HEL} like weights, invariants and Lie algebras
are translated into the language of residual re\-pre\-sen\-ta\-tions.

In analogy to Lie groups like the compact $\U(n)$ or
the noncompact $\D(1)$ also  
symmetric spaces like 
the noncompact position   manifold $\SD(2)$ and  spacetime $\D(2)$ 
have linear re\-pre\-sen\-ta\-tions which
will be considered 
in analogy to
the re\-pre\-sen\-ta\-tions of the time group $\D(1)$.
To construct residual re\-pre\-sen\-ta\-tions 
of  the rank 2  spacetime manifold $\D(2)$ 
distributions 
of the  energy-mo\-men\-ta  $q\in \R^4$ (tangent space forms) 
are used, supported by  two invariant masses $q^2\in\{m^2_0,m^2_3\}$
characterizing the Cartan subgroup $\D(\bl1_2)\x\SO_0(1,1)$-re\-pre\-sen\-ta\-tions
for time and position.

\section{Quantum Representations of Time}

A dynamics is a re\-pre\-sen\-ta\-tion of time, realized in quantum mechanics
by the
quantization  (anti-) commutators of the quantum algebra generating operators. 
In the simplest cases of a harmonic oscillator 
with Hamiltonian
$H={\bl p^2\over 2M}+m^2M{\bl x^2\over2}$ 
for mass $M$ and frequency $m$ or of a free mass point
with $H={\bl p^2\over 2M}$
for  frequency $m\to 0$  the time dependent commutation relations of the 
dual quantum algebra generating position-momentum pair 
$(\bl x,\bl p)$ 
give the time re\-pre\-sen\-ta\-tion matrix elements
\begin{eq}{l}
\D(1)\ni e^t\mape D(t)={\scriptsize\pmatrix{
\com{i\bl p}{\bl x}&\com {\bl x}{\bl x}\cr
\com {\bl p}{\bl p}&\com {\bl x}{-i\bl p}\cr}}(t)=\left\{
\begin{array}{cl}
{\scriptsize\pmatrix{
\cos t m &{i\over M m }\sin t m \cr 
iM m  \sin t m &\cos t m \cr}}&\in\SO(2)\cr
{\scriptsize\pmatrix{
1&{it\over M}\cr
0&1\cr}}&\in\U(1,1)\end{array}\right.
\end{eq}with the shorthand notation
$\com{a(s)}{b(t)}_\ep=\com ab_\ep(t-s)$, $\ep=\pm1$,
 valid for all matrix elements.
Those re\-pre\-sen\-ta\-tions 
 arise from
the complex  irreducible and nondecomposable  time re\-pre\-sen\-ta\-tions
with creation and annihilation operator $(\ro u,\ro u^\star)$
and nil- and eigenoperators\cite{S89}
 $(\ro b,\ro g,\ro b^\x,\ro g^\x)$  resp.
\begin{eq}{l}
\D(1)\ni e ^t\mape \left\{
\begin{array}{cll}
\com{\ro u^\star}{\ro u}_\ep(t)&= e ^{ti m}&\in\U(1)\cr
{\scriptsize\pmatrix{
\com{\ro g^\x}{\ro b}_\ep&  \com{\ro b^\x}{\ro b}_\ep\cr
\com{\ro g^\x}{\ro g}_\ep&  \com{\ro b^\x}{\ro g}_\ep\cr}}(t)&=
{\scriptsize\pmatrix{1& ti\nu \cr 0&1\cr}} e^{tim  }
&\in \U(1,1)\end{array}\right.
\end{eq}

The quantization opposite commutators implement
the Lie algebra of the basic space endomorphisms, e.g.
the Hamiltonians above.
For the harmonic oscillator 
the $\U(1)$-induced Fock form $\Ffo{...}$ of the time dependent anticommutators
arises as time derivative of the quantization
\begin{eq}{l}
{\scriptsize\pmatrix{
\Ffo{\acom{i\bl p}{\bl x}}&\Ffo{\acom {\bl x}{\bl x}}\cr
\Ffo{\acom {\bl p}{\bl p}}&\Ffo{\acom {\bl x}{-i\bl p}}\cr}}(t)
= {\scriptsize\pmatrix{
i\sin t m &{1\over M m }\cos t m \cr 
M m  \cos t m &i\sin t m \cr}}={1\over im}{d\over dt}D(t)
\end{eq}

For the general quantum mechanical case with 
$iH=i[{\bl p^2\over 2M}+V(\bl x)]$
as basis for the represented Lie algebra\footnote{\scriptsize 
The Lie group to Lie algebra transition
$G\mape \log G$ is denoted with the logarithm $\log$
as covariant functor.}
 $\log\D(1)\cong\R$
the time $\D(1)$-re\-pre\-sen\-ta\-tion 
matrix elements as the ground state values 
$\angle{\com{a(s)}{b(t)}_\ep}=\angle{\com ab_\ep}(t-s)$
of the position-momentum commutators 
can be computed from  the imaginary and time translation antisymmetric
position commutator
\begin{eq}{l}
\angle{\com {\bl x}{\bl x}}(t)=
\int_0^\infty  d m^2\mu(m^2)
i{\sin tm\over Mm}
\end{eq}with a spectral measure $\mu(m^2)$
for the time translation eigenvalues $m\in\R$ (frequencies, energies),
e.g. $\mu(m^2)=\de(m^2-m_0^2)$
with $m_0^2>0$ for oscillator and $m_0=0$ for free mass point,
and $\bl p=M{d\bl x\over dt}$
\begin{eq}{l}
\angle{{\scriptsize\pmatrix{
\com{i\bl p}{\bl x}&\com {\bl x}{\bl x}\cr
\com {\bl p}{\bl p}&\com {\bl x}{-i\bl p}\cr}}   }(t)=
\int_0^\infty  d m^2\mu(m^2){\scriptsize\pmatrix{
\cos tm&{i\over Mm}\sin tm\cr
iMm \sin tm&\cos tm\cr}}\in\rep\SO(2)
\end{eq}In the case of a compact time development,
i.e. representations in $\U(1)$ or $\SO(2)$, where there 
exists  a basis of normalizable energy eigenvectors (for the oscillator
build by the monomials of creation and annihilation operator),
the energy measure is  definite  $\mu(m^2)\ge0$.

\section{Time and Position Translations}

\subsection{The Lie Groups for the Translations}

Translations are formalized by additive groups (vector spaces) $\R^n$. It will be convenient
to introduce a distinguishing notation for the Lie group and the Lie
algebra
involved which have an isomorphic Abelian Lie group structure
\begin{eq}{l}
\left.\begin{array}{llll}
\hbox{Lie group}&\D(1)&=\exp\R&=\{e^x\mid x\in\R\}\cr
\hbox{Lie algebra}&\R&=\log\D(1)&=\{x\mid x\in\R\}\cr\end{array}\right\},~~
\exp \R\cong\R
\end{eq}The noncompact  group $\D(1)$ as universal covering  group
is locally isomorphic
to the compact one $e^{i\al}\in \U(1)=\exp i\R\cong\R/\Z$ with 
 Lie algebra
$\log\U(1)=i\R$.

The  groups $\U(1)$ and $\D(1)$
are - as real 1-dimensional Lie groups - isomorphic to the axial rotations $\SO(2)$
and the Procrustes\footnote{\scriptsize
Procrustes in the Greek mythology either shrinked or stretched
his visitors - tall or short resp. - to death.}
 dilatation  group  $\SO_0(1,1)$ resp., i.e.
the 1-dimensional boosts 
\begin{eq}{rlll}
\hbox{compact}&\U(1)&\cong\SO(2)&=\{
{\scriptsize\pmatrix{\cos \al&\sin \al\cr-\sin \al &\cos\al\cr}}\mid\al\in\R\}\cr
\hbox{noncompact}&\D(1)&\cong\SO_0(1,1)&=\{
{\scriptsize\pmatrix{\cosh x&\sinh x\cr\sinh x &\cosh x\cr}}\mid x\in\R\}\cr
\end{eq}Those orthogonal groups with
invariant bilinear forms of the 2-dimensional 
vector space they are acting upon,
will be called selfdual 
representations\footnote{\scriptsize
For a group and a Lie algebra dual representations 
on finite dimensional dual vector spaces are
related to each other by inverse and negative transposition resp.}
 of $\U(1)$ and $\D(1)$ resp. with the obvious
isomorphy (for $\SO(2)$ only in the complex)
\begin{eq}{rllll}
\hbox{definite unitary:}&
\SO(2)&\ni
{\scriptsize\pmatrix{\cos \al&i\sin \al\cr i\sin \al &\cos\al\cr}}
&\cong {\scriptsize\pmatrix{e^{i\al}&0\cr0&e^{-i\al}\cr}}
&\in\SU(2)\cr
\hbox{indefinite unitary:}&
\SO_0(1,1)&\ni
{\scriptsize\pmatrix{\cosh x&\sinh x\cr\sinh x &\cosh x\cr}}
&\cong {\scriptsize\pmatrix{e^x&0\cr 0 &e^{-x}\cr}}
&\in\SU(1,1)
\end{eq}

\subsection{Real Operations have Unitary Representations} 

The algebraic and topological completeness 
of the complex field $\C$ allows the
definition of the transcendental number $e$ involving
`exponential completeness'
$\exp\C=\C\setminus\{0\}$ and, therewith, the
exponential transition from local
linear structures (tangent vector spaces, Lie algebras) to global
possibly nonlinear structures 
(symmetric spaces, Lie groups). Therefore, 
I will consider representations on complex
vector spaces only. The complex representations of the
physically arising only real Lie groups or Lie algebras have 
to be unitary - definite $\U(n)$ or
indefinite $\U(p,q)$, in order to recognize the realness  also in the 
representation. Therewith, the complex numbers are always used together
with the canonical conjugation, i.e. as the doubled real field 
$\C=\R\pl i\R$. 

Only for one complex dimension unitarity is unique, characterized by
the real Lie group $\U(1)=\exp i\R$. 
The $n$ unitarities for $n$ complex dimensions go with the signature:
E.g., in two dimensions 
the $\U(2)$-conjugation
of $2\x2$-matrices can be written as the familiar conjugate transposition
which exchanges the elements of the skewdiagonal 
whereas the $\U(1,1)$-conjugation can be written
with an exchange  of the diagonal elements
\begin{eq}{rl}
\hbox{$\U(2)$-conjugation:}&
{\scriptsize\pmatrix{\al&\be\cr\ga&\de\cr}}
\lrmap 
{\scriptsize\pmatrix{\ol\al&\ol\ga\cr\ol\be&\ol\de\cr}}\cr
\hbox{$\U(1,1)$-conjugation:}&
{\scriptsize\pmatrix{\al&\be\cr\ga&\de\cr}}
\lrmap 
{\scriptsize\pmatrix{\ol\de&\ol\be\cr\ol\ga&\ol\al\cr}}
={\scriptsize\pmatrix{0&1\cr1&0\cr}}
{\scriptsize\pmatrix{\ol\al&\ol\ga\cr\ol\be&\ol\de\cr}}
{\scriptsize\pmatrix{0&1\cr1&0\cr}}\cr
\end{eq}

\subsection{Nildimensions for Noncompact  Groups}

 Noncompact groups have reducible, but nondecomposable
representations\cite{BOE,S89,SHELO}
where  the representation space cannot be spanned by
eigenvectors only - 
there occur also nilvectors, i.e. principal vectors which are not
eigenvectors. The linear operators involved have a Jordan
 triangular form
with nontrivial off-diagonal entries.

The situation is characterized by the nondecomposable representations of the
group $\D(1)$ with an eigenvalue $m$ for 
$e^x\mape e^{xim}$ which comes 
multiplied with an automorphism of the representation space
$V\cong\C^{1+N}$ and can be written with a nilcyclic matrix $M_N$
(nil-Hamiltonian),
nilpotent to the power $N+1$
\begin{eq}{r}
\D(1)\ni e^x\mape e^{xi(m+ M_N)}
\cong e^{xim}{\scriptsize\pmatrix{
1&ix&{(ix)^2\over 2!}&\dots&{(ix)^N\over N!}\cr 
0&1&ix&\dots&{(ix)^{N-1}\over (N-1)!}\cr 
&\dots&&&\dots\cr
0&\dots&0&1&ix\cr
0&\dots&0&0&1\cr}}\in\GL(\C^{1+N})\cr
(M_N)^N\ne0,~~(M_N)^{N+1}=0,~~
M_N\cong{\scriptsize\pmatrix{
0&1&0&\dots&0\cr
0&0&1&\dots&0\cr
&\dots&&\dots&\cr
0&0&\dots&0&1\cr
0&0&\dots&0&0\cr}}
\end{eq}The natural number 
$N$ is called the nildimension with $1+N$
the dimension of the nondecomposable representation.
Irreducible representations have
trivial nildimension $N=0$ and  $M_0=0$. 
For $N\ge1$ the conjugation is indefinite, 
i.e. the group image is a subgroup of $\U(1,1),\U(2,1),\U(2,2)$ etc.

An example for nontrivial nildimensions in quantum mechanics is
the radial part $\psi_{nL}$ of the bound state wave functions
in the hydrogen atom: It  is a linear combination
of matrix elements $r\mape r^N e^{-{r\over k}}$
 of noncompact representations of the radial
translations with eigenvalue $-{1\over k}$, $k=n+L+1$
\begin{eq}{ll}
\R^+\ni r\mape D_{nL}(r)= r\psi_{nL}(r)\sim
 \({2r\over k}\)^{L+1}\cl L^{2L+1}_n({2r\over k})~e^{-{r\over k}}
\end{eq}with the Laguerre polynomials 
$\cl L$ as combinations of radial powers $r^N$.

An example for nontrivial nildimensions in quantum field theory
is quantum electrodynamics where the nonparticle components of 
the $\U(1)$-gauge field which come 
in addition to the left and right
circularily polarized particle degrees of freedom (photons),
i.e. the Coulomb force inducing degree of freedom and
the so called gauge degree of freedom, are 
spacetime translation nilvectors\cite{S923,SBH95}, 
i.e. principal vectors
 which are no eigenvectors.  
The dichotomy between particles 
and interaction degrees of freedom
in the electromagnetic potential reflects
 the compact and noncompact Cartan subgroups
in the Lorentz group  $\SO(2)\x\SO_0(1,1)\subnoteq\SO_0(1,3)$,
represented definite unitarily $\SO(2)\map\U(2)$ for the photons 
and indefinite unitarily $\SO_0(1,1)\map\U(1,1)$
for Coulomb and gauge degree of freedom.
The  nilpotency of the BRS-generator\cite{BRS} with power $N+1=2$
has its origin in the time translation representation 
$\D(1)\map\U(1,1)$  for the
two nonparticle degrees of freedom with 
$M_1={\scriptsize\pmatrix{0&1\cr0&0\cr}}$ the nil-Hamiltonian
which fulfills  $M_1^2=0$.

\section
{The Spacetime Representation Structure of Quantum Particle Fields}

Particle fields are appropriate to describe {free} particles,
they implement definite unitary re\-pre\-sen\-ta\-tions of 
the Poincar\'e Lie algebra\cite{WIG,MACK} 
$\log \SO_0(1,3)\rvec\pl \R^4$.

A particle field, in the simplest case a hermitian 
scalar massive field
$\bl \Phi$, $m>0$, with creation and annihilation operators
$(\ro u,\ro u^\star)$
\begin{eq}{l}
\bl\Phi(x)
=\int{d^3 q\over (2\pi)^3}\sqrt{{m\over q_0}}~ 
{  e^{xiq}\ro u(\rvec q)+
 e^{-xiq}\ro u^\star(\rvec q)\over\sqrt2},~~q_0=\sqrt{m^2+\rvec q^2}\cr
\com{ \ro u^\star(\rvec p)}{\ro u (\rvec q)}=(2\pi)^3 \de(\rvec q-\rvec p)
=\angle{\acom{ \ro u^\star(\rvec p)}{\ro u (\rvec q)}}
=\angle{\ro u^\star(\rvec p)\ro u (\rvec q)}\cr
\end{eq}is characterized by its quantization\footnote{\scriptsize
The linear Minkowski   spacetime parametrization is used
in the  notation for (anti)commutators
$[A(y),B(x)]_\pm=[A,B]_\pm(x-y)$.}, causally supported and on shell
\begin{eq}{l}
{\com{\bl\Phi}{\bl\Phi}(x)\over m}=i{\bl s(x|m)\over m}
=\int{d^4q\over(2\pi)^3}\ep(q_0)\de(q^2-m^2) e^{xiq}=0\hbox{ for }x^2<0\cr
\end{eq}and its Feynman propagator adding up the 
Fock form value of the quantization-opposite commutator, also on shell
\begin{eq}{l}
{\Ffo{\acom{\bl\Phi}{\bl\Phi}}(x)\over m}={\bl C(x|m)\over m}
=\int{d^4q\over(2\pi)^3}\de(q^2-m^2) e^{xiq}\cr
\end{eq}and
the $\ep(x_0)$-multiplied quantization\cite{GELSHIL1} 
which has also off shell
contributions, i.e. for  $q^2\ne m^2$
\begin{eq}{rl}
{\ep(x_0)\bl s(x|m)\over m}
&={1\over \pi}\int{d^4q\over(2\pi)^3}{1\over -q_\ro P^2+m^2}
e^{xiq}\hbox{ (principal value $\ro P$)}\cr
{\Ffo{\acom{\bl\Phi}{\bl\Phi}(x)
\pm \ep(x_0)\com{\bl\Phi}{\bl\Phi}(x)}\over m}
&={\bl C(x|m)\pm i\ep(x_0)\bl s(x|m)\over m}
=\pm{1\over i\pi}\int{d^4q\over(2\pi)^3}{1\over q^2\mp io-m^2}
e^{xiq}\cr
\end{eq}The harmonic contributions in the quantization 
\begin{eq}{l}
i{\bl s(x|m)\over m}
=\int{dq_0\over(2\pi)^2}e^{x_0iq_0}
\ep(q_0)\vth(q_0^2-m^2){\sin r \sqrt{q_0^2-m^2}\over r }\cr
\end{eq}and in the Feynman propagator
\begin{eq}{l}
{\bl C(x|m)\over m}\pm {\ep(x_0) i \bl s(x|m)\over m}\cr
=\int{dq_0\over(2\pi)^2}e^{x_0iq_0}\left\{\begin{array}{l}
\hskip1mm\bigl[\vth(q_0^2-m^2){\sin r \sqrt{q_0^2-m^2}\over r }\cr
\hskip-2mm\pm i\vth(q_0^2-m^2){\cos r \sqrt{q_0^2-m^2}\over r }
\pm i\vth(m^2-q_0^2){e^{- r \sqrt{m^2-q_0^2}}\over r } \bigr]\end{array}\right.\cr
\end{eq}show irreducible (definite unitary) time translation  representation 
matrix elements
\begin{eq}{l}
\R\ni x_0\mape e^{\pm x_0iq_0}\in\U(1)
\end{eq}With the polar coordinate
position translation decomposition 
\begin{eq}{l}
\rvec x\in\R^3\cong\R^+\x\SO(3)/\SO(2)
\end{eq}and the
geometrical Kepler factor
${1\over r}$ for the sphere surface $\SO(3)/\SO(2)$-distribution,
the  position radial  translation monoid 
$r\in\R^+$ is   represented 
by ${\sin r|\rvec q|\over r}$ (spherical Bessel function)
with $\sin r|\rvec q|$ as matrix element of a compact  group
for  the 
quantization $\bl s(x|m)$ and the Fock form function $\bl C(x|m)$.
In the propagator contribution
$\ep(x_0)\bl s(x|m)$ there arise the $r=0$-singular spherical
Neumann function ${\cos r|\rvec q|\over r}$ which contains 
$\cos r|\rvec q|$ as a compact position translation 
representation matrix element. 
The additional off shell
induced Yukawa contributions displays
a  representation matrix element of the
radial position translations in a noncompact (indefinite unitary) group
\begin{eq}{l}
\R^+\ni r\mape\left\{\begin{array}{ll} 
e^{\pm ri|\rvec q|}&\in\SO(2)\cr
e^{-r|Q|}&\in\SO_0(1,1)\cr\end{array}\right.
\end{eq}The off shell contributions with the Yukawa interactions 
in the Feynman propagator
are no definite unitary Poincar\'e Lie algebra representation matrix elements.

The time projection $\int d^3 x$ of quantization and Feynman propagator
gives matrix elements for the representation of 
time translations  in the rest system   of a massive particle 
\begin{eq}{l}
x_0\mape \int d^3 x
{\scriptsize\pmatrix{
\bl C(x|m)\cr
 i\bl s(x|m)\cr
\ep(x_0)i\bl s(x|m)\cr}}=
 \int dE\ep(E){\scriptsize\pmatrix{E\cr m\cr m\ep(x_0)\cr}}
\de(E^2-m^2)e^{x_0iE}
={\scriptsize\pmatrix{
\cos x_0 m\cr
i\sin x_0m\cr
i\sin |x_0|m\cr}}
\end{eq}The analogue position projection $\int dx_0$
\begin{eq}{ll}
\rvec x\mape 2\pi\int {d x_0\over im} {\scriptsize\pmatrix{
i\bl s(x|m)\cr
\bl C(x|m)\cr
\ep(x_0)i\bl s(x|m)\cr}}=
\int{dQ\over 2 }{\scriptsize\pmatrix{0\cr 0\cr
\vth(Q^2-m^2)\cr}}
e^{-r|Q|}={\scriptsize\pmatrix{0\cr0\cr
{e^{-rm}\over  r}\cr}}
\end{eq}is nontrivial only for the off shell contributions with 
radial translation representation matrix element $e^{-rm}$ in a noncompact
group. 

Particle fields display 
in the quantization $\bl s(x|m)$ and the
Fock form $\bl C(x|m)$, both on shell $q^2=m^2$,
matrix elements of  definite unitary  representations for the 
translations.
The off shell contributions in  $\ep(x_0)\bl s(x|m)$
involve  matrix elements for indefinite unitary representation matrix elements
for position translations $\R^3$.

\section{Homogeneous Models for Time, Position and  Spacetime}

\subsection{Exponentiating Time Translations}

The time translations as a real 1-dimensional vector space 
$x_0=\ol{x_0}\in\R$
are isomorphic - as Lie group - to its exponent
$\D(1)=\exp\R$, the time group. 
They constitute  the noncompact part (modulus) of the full complex group,
given by the phase classes
\begin{eq}{l}
\hbox{time: }\GL(\C)/\U(1)=\D(1)=\exp\R\cong\R
\end{eq}

\subsection{Exponentiating Position Translations}

In the semidirect Euclidean group
$\SO(3)\sx\R^3$
the position translations as a real 3-dimensional vector space $\R^3$
are isomorphic - as vector space with rotation action - 
to the Lie algebra of the rotations $\log\SO(3)\cong\R^3$.
In the $\SU(2)$-formulation, the rotations $\SO(3)$ are representated
by the  adjoint action of its covering  group $\SU(2)$ 
\begin{eq}{l}
\SO(3)\sx\R^3\sim\SU(2)\sx\R^3,~~O.\rvec x\sim u\o \rvec x\rvec \si \o u^{-1}\cr
\hbox{with }\rvec x \rvec\si
={\scriptsize\pmatrix{x_3&x_1-ix_2\cr x_1+ix_2&-x_3\cr}}\cr
u\in\SU(2)\then
O^a_b={1\over2}\tr\si^au\si^bu^{-1},~~O\in\SO(3)\cong\SU(2)/\{\pm\bl1_2\}
\end{eq}In the Pauli representation, the position translations are
hermitian $2\x2$-matrices, i.e. representatives\footnote{\scriptsize
The funny double element symbol
means a representative of a coset, i.e.
$g\rin G/H\iff g\in gH\in G/H$.}  of the
classes of all complex special matrices 
$\log\SL(\C^2)\cong \R^3\pl(i\R)^3$ with respect to the 
special unitary ones $\log \SU(2)\cong (i\R)^3$
\begin{eq}{l}
\rvec x\rvec \si=(\rvec x\rvec \si)^\star\rin \log\SL(\C^2)/\log\SU(2)
\end{eq}The global position manifold arises by exponentation, isomorphic
as symmetric space to the
classes of the Lorentz covering group $\SL(\C^2)$ with respect to
the rotation covering group $\SU(2)$ 
\begin{eq}{l}
\hbox{position: }\SL(\C^2)/\SU(2)\cong\SD(2)=\exp\R^3\cong\R^3
\end{eq}The  global symmetric space position $\SD(2)$ and its tangent 
vector space $\R^3$
have a  manifold isomorphy only, $\exp \R^3\ne (\exp \R)^3$.

\subsection{Exponentiating Spacetime Translations}

In the  Poincar\'e  group
$\SO_0(1,3)\sx\R^4$
the translations  $\R^4$
are not isomorphic
to the Lie algebra of the Lorentz group  $\log\SO_0(1,3)\cong\R^6$.
In the $\SL(\C^2)$-formulation, the Lorentz transformations
$\SO_0(1,3)$ are represented by the  conjugate adjoint action of 
its covering group $\SL(\C^2)$
\begin{eq}{l}
\SO_0(1,3)\sx\R^4\sim\SL(\C^2)\sx\R^4,~~\La.x\sim s\o x\o s^\star\cr
\hbox{with }x= x_k\si^k
={\scriptsize\pmatrix{x_0+x_3&x_1-ix_2\cr x_1+ix_2&x_0-x_3\cr}}\cr
s\in\SL(\C^2)\then \La^k_j={1\over2}\tr\si^k s\d \si_j s^\star,~~
\La\in\SO_0(1,3)\cong\SL(\C^2)/\{\pm\bl1_2\}
\end{eq}with Weyl matrices 
$\si^k=(\bl1_2,\rvec \si)=\d\si_k$.
In the Cartan  representation, the spacetime translations are
 hermitian $2\x2$-matrices, i.e. representatives
of the
classes of all complex  matrices 
$\log\GL(\C^2)\cong \R^4\pl(i\R)^4$ with respect to the 
unitary ones $\log \U(2)\cong (i\R)^4$
\begin{eq}{l}
 x= x^\star\rin \log\GL(\C^2)/\log\U(2)
\end{eq}Global spacetime arises by exponentation and is given by the
classes of the full group $\GL(\C^2)$ with respect to
the unitary phases $\U(2)$, the moduli of $\GL(\C^2)$ 
\begin{eq}{l}
\hbox{spacetime: }\GL(\C^2)/\U(2)\cong \D(2)=\exp\R^4\cong\R^4
\end{eq}The causal structure of spacetime is the 
spectral order\cite{RIC} of the $C*$-algebra $\log\GL(\C^2)$.

The noncompact symmetric space $\D(2)$ 
 has - analogue to
its compact counterpart
$\U(2)$ with $\U(2)=\U(\bl1_2)\o\SU(2)$ - a product decomposition into
Abelian  causal time group $\D(\bl1_2)$ 
and real 3-dimensional position (boost) manifold $\SD(2)$
\begin{eq}{l}
\D(2)=\D(\bl1_2)\x\SD(2),~~\SD(2)\cong\SL(\C^2)/\SU(2)
\end{eq}Both symmetric spaces have real rank $2$ 
- also indicated  in  the notation $\U(2)$ and $
\D(2)$ -  which reflects both the number of independent invariants
and 
the
dimension of a maximal Abelian  Cartan subgroup (flat submanifold\cite{HEL}), 
arising as factor 
of the 2-sphere $\SO(3)/\SO(2)$
in the polar decomposition
\begin{eq}{lclclcl}
\U(2)&=&\U(\bl 1_2)\o\SU(2)&\cong&\U(1)\o\SO(2)&\x&\SO(3)/\SO(2)\cr
\D(2)&=&\D(\bl1_2)\x\SD(2)&\cong&\D(1)\x\SO_0(1,1)&\x&\SO(3)/\SO(2)\cr
\end{eq}For the decomposition of the 
real 4-dimensional tangent spaces (Lie algebra for $\U(2)$)
with the Lie algebra of the Cartan subgroup the sphere
factor remains  unchanged
\begin{eq}{ll}
\log\U(2)&=\log\U(\bl 1_2)\pl\log\SU(2)\cr
&\cong
\log\U(1)\pl[ \log\SO(2)\x\SO(3)/\SO(2)]\cr
\log\D(2)&=\log\D(\bl1_2)\pl\log \SD(2)\cr
&\cong \log \D(1)\pl[\log \SO_0(1,1)\x\SO(3)/\SO(2)]\cr
\end{eq}

The representations of noncompact
spacetime $\D(2)$ and compact internal group $\U(2)$ are characterized
by two invariants 
from a continuous spectrum for a Cartan 
subgroup $\D(1)\x\SO_0(1,1)$ and
from a discrete spectrum for a Cartan 
subgroup $\U(1)\o\SO(2)$ resp.
Minkowski spacetime $\R^4$ in the Cartan representation by $\U(2)$-hermitian
$2\x2$-matrices has the  familiar conjugate adjoint $\GL(\C^2)$-transformation
behaviour to be compared with the adjoint action of the compact group
$\U(2)$ on its Lie algebra $\log\U(2)\cong(i\R)^4$
\begin{eq}{lllll}
g\in\GL(\C^2),&
x&={\scriptsize\pmatrix{x_0+x_3&x_1-ix_2\cr x_1+ix_2&x_0-x_3\cr}}&\in\log\D(2)
&\then x\mape g\o x\o g^*\cr
u\in\U(2),&
i\al&=
i{\scriptsize\pmatrix{\al_0+\al_3&\al_1-i\al_2\cr \al_1+i\al_2&\al_0-\al_3\cr}}
&\in\log\U(2)
&\then i\al\mape u\o i\al\o u^*\cr
&&&&\hskip10mm
u^*=u^{-1}\cr
\end{eq}However, in contrast to the decomposition of the $\U(2)$-Lie
algebra into Abelian $\U(\bl1_2)$ and simple $\SU(2)$-contribution,
compatible with the adjoint $\U(2)$-action,  the decomposition
of spacetime $\D(2)$ and its tangent space
into time and position is not compatible with the action
of the Lorentz  group
\begin{eq}{lll} 
u\in\U(2),&\log\U(2)\ni i\al=i\al_0\bl1_2+i\rvec\al\rvec\si,&
\left\{\begin{array}{ll}
u\o i\al_0\bl1_2\o u^*&\in\log \U(\bl1_2)\cr
u\o i\rvec\al\rvec\si\o u^*&\in\log\SU(2)\cr\end{array}\right.\cr
s\in\SL(\C^2),&\log\D(2)\ni x=x_0\bl1_2+\rvec x\rvec\si&\cr
&\hfill\hbox{in general}&\left\{\begin{array}{ll}
s\o x_0\bl 1_2\o s^*&\notin\log \D(\bl1_2)\cr
s\o \rvec x\rvec\si\o s^*&\notin\log\SD(2)\cr\end{array}\right.\cr
\end{eq}

Both symmetric spaces are parametrizable by exponentiating the
tangent space, e.g. in the polar Cartan decomposition
\begin{eq}{rll}
\log\U(2)&\ni i\al
&=u({\rvec\al\over|\rvec\al|})
\o i(\al_0\bl1_2+|\rvec \al|\si_3)\o u^*({\rvec\al\over|\rvec\al|})\cr
\then &\exp i\al 
&=u({\rvec\al\over|\rvec\al|}) \o e^{i(\al_0\bl1_2+|\rvec \al|\si_3)}\o 
u^*({\rvec\al\over|\rvec\al|})\in\U(2)\cr
\log\D(2)&\ni x
&=u({\rvec x\over  r})\o 
(x_0\bl1_2+ r \si_3)\o u^*({\rvec x\over r }),~~ r= |\rvec x| \cr
\then& \exp x 
&=u({\rvec x\over r })\o 
e^{x_0\bl1_2+ r \si_3}\o 
u^*({\rvec x\over r })\in\D(2)\cr
\end{eq}The  diagonalization of
 $\D(2)$ and $\U(2)$ with the sphere operations
\begin{eq}{l}
u({\rvec x\over r })={\scriptsize\pmatrix{
\cos{\th\over2}&-e^{-i\phi}\sin{\th\over2}\cr
e^{i\phi}\sin{\th\over2}&\cos{\th\over2}\cr}}
\rin\SU(2)/\U(1)\cong\SO(3)/\SO(2)
\end{eq}defines 
$\{i\al_0,i|\rvec\al|\}$ as Cartan coordinates for
the internal group  
and  $\{x_0, r \}$ (time and radial translations) 
as Cartan coordinates for spacetime.

Similar to the local-global 
group isomorphism for time $\R\cong\exp\R=\D(1)$ one has
the manifold isomorphy  for spacetime $\R^4\cong\exp\R^4=\D(2)$. 
Via their embedding as future cones $\D(1)$ and $\D(2)$
are parametrizable with tangent space $\R$ and $\R^4$ coordinates 
\begin{eq}{llllll}
t\in\R&\hskip-3mm\then&\hskip-3mm\D(1)\ni e^t=\ep(s) s&\in\R^+
&\hskip-2mm\hbox{with }s\in\R,&s^2=e^{2t}\cr
x\in\R^4&\hskip-3mm\then &\hskip-3mm\D(2)\ni e^x=\ep(y_0)\vth(y^2)y
&\in(\R^4)^+ &\hskip-2mm\hbox{with }y\in\R^4,&
\left\{\begin{array}{ll}
y_0^2&=e^{2x_0}\cr
|\rvec y|&=e^{ r }\end{array}\right.\cr
\end{eq}

\subsection{Time in Spacetime}

A dynamics  in quantum mechanics 
arises from representations of the time group $\D(1)\cong\exp \R$
whose representation spaces  are realized
in the Schr\"odinger picture 
by wave functions depending on position translations.
The quantum mechanical relevant time structure 
is a  proper substructure of spacetime,  modeled
by the homogeneous  space $\D(2)\cong\GL(\C^2)/\U(2)$
and represented by  quantum fields.
The quantum mechanical energy eigenstates
for compact $\D(1)$-representations  are embedded as
spacetime  particles.
The strict  future cone  with dimension four in flat spacetime being
isomorphic to nonlinear spacetime
$\D(2)$ 
contains not only the totally ordered 
1-di\-men\-sio\-nal causal subgroup $\D(1)$, it leaves room
for a 3-di\-men\-sio\-nal position  submanifold $\SD(2)$ whose
noncompact dilatations $\SO_0(1,1)$ characterize
spacetime interactions.   
The 
particle contributions, unitarily representing $\D(1)$,
 have to be
supplemented in relativistic quantum theories
by nonparticle ones to implement genuine 
$\SO_0(1,1)$-re\-pre\-sen\-ta\-tions.
The nonparticle contributions  are a
genuine intrinsic feature of spacetime $\D(2)$ 
without  analogue in quantum mechanics.
There the interactions, e.g. the Coulomb potential for atoms,
have to be put in by hand

\begin{eq}{c}\hskip0mm
\begin{array}{|c||c|c|c|}\hline
&\begin{array}{c}
\hbox{time}\cr
\D(1)=\GL(\C)/\U(1)\end{array}
&\inmap&\begin{array}{c}
\hbox{spacetime}\cr
\D(2)\cong\GL(\C^2)/\U(2)\end{array}\cr\hline\hline
\hbox{quantum theory}&
\hbox{quantum mechanics}&\inmap&
\hbox{quantum fields}\cr\hline
\hbox{Cartan subgroup}&\D(1)&\inmap&\D(1)\x\SO_0(1,1)\cr\hline
\hbox{full group}&\GL(\C)&\inmap&\GL(\C^2)\cr\hline
\begin{array}{c}
\hbox{tangent space}\cr
\hbox{(translations)}\cr\end{array}
&\R&\inmap&\R^4\cr\hline
\hbox{future}&t=\ep(t)t&\inmap 
&x=\ep(x_0)\vth(x^2)x\cr
&\R^+\cong\D(1)&&(\R^4)^+\cong\D(2)\cr\hline
\begin{array}{c}
\hbox{particles}\cr
\hbox{(states)}\end{array}&\D(1)\map \U(1)&\cong&
\D(1)\map\U(1)\cr\hline
\hbox{interactions}&\hbox{not intrinsic}&&
\SO_0(1,1)\map\U(1,1)\cr\hline
\end{array}
\end{eq}

\section
{Two Continuous Invariants for Spacetime Representations}

Since Yukawa, the unification of a 
causal time development, characterized by a
particle mass $ m _0\ge0$ with a position   interaction,
characterized by a  range ${1\over m _3}$, $m_3\ge0$,
in one spacetime Klein-Gordon equation for an $\ep(x_0)$-multiplied 
quantization distribution with one mass $m\ge 0$
\begin{eq}{r}
\left.\begin{array}{rll}
({d^2\over dt^2}+ m _0^2)&{  e ^{|t|im_0}\over im_0}&=2\de(t)\cr
(-{\p^2\over\p\rvec x^2}+ m _3^2)&{  e ^{- r m_3}\over2\pi r }&=
2\de(\rvec x)\cr\end{array}\right\}\inmap
(\p^2+ m ^2)\ep(x_0){\bl s(x|m)\over m}=2\de(x)\cr
\hbox{ with } m_0= m_3= m 
\end{eq}seems to be an obvious relativistic bonus -
all interactions can be interpreted as particle induced.

Particle fields with a Dirac energy-mo\-men\-tum distribution 
in their quantization
\begin{eq}{l}
i\bl s(x|m)
=\int {d^4q\over(2\pi)^3}\ep(q_0)
m\de(q^2- m^2)
  e ^{xiq}\cr
\end{eq}give by  position  integration 
re\-pre\-sen\-ta\-tion matrix elements of the 
Abelian time group $\D(1)\cong\exp\R$ in $\SO(2)$
\begin{eq}{rl}
\D(1)&\map\C\cr
e^{x_0}&\mape\int d^3 x~ i\bl s (x|m)
 =\int dE~ m\ep(E) \de(E^2- m^2)~  e ^{ x_0 iE}=i\sin x_0m\cr
\end{eq}

The appropriate  distribution for a re\-pre\-sen\-ta\-tion 
of the position symmetric space $\SD(2)\cong\exp\R^3$  arises from
a derived energy-mo\-men\-tum Dirac distribution 
\begin{eq}{l}
{i\bl s^{\rm dip}(x| m)\over m}=
-{d\over d m^2}
{i\bl s(x| m)\over m}=
\int {d^4q\over(2\pi)^3}\ep(q_0)
 \de'(q^2- m^2)
  e ^{xiq}\cr
\end{eq}Time integration leads to
a Dirac distribution for the invariant
and to $\SD(2)$-re\-pre\-sen\-ta\-tion matrix elements
in $\SO_0(1,1)$
\begin{eq}{rl}
\SD(2)&\map\C\cr
e^{-\rvec x}&\mape4\pi\int dx_0\ep(x_0)
\bl s^{\rm dip}
(x| m)
=\int dQ~ m
\de(Q^2-  m^2)~  e ^{- r|Q|}=e ^{- r  m}\cr
\end{eq}

The  Dirac energy-momentum distribution 
for time with characterizing 2nd order differential equation 
in contrast to the  derived  distribution 
for position with  characterizing  4th order differential equation
\begin{eq}{l}
({d^2\over dt^2}+ m^2){  e ^{|t|im}\over im}=2\de(t),~~
(-{\p^2\over\p\rvec x^2}+  m^2)^2
{  e ^{- r m}\over 4\pi  m}=2\de(\rvec x)
\end{eq}reflect the different dimensions 1 and 3 
of the time group $\D(1)$ and the position   manifold $\SD(2)$
resp.

The association of 
energy-momentum  singularities 
to  re\-pre\-sen\-ta\-tion invariants for $\D(1)$ (time) and $\SD(2)$ 
(position)
resp. is blurred since a decomposition
 of the spacetime tangent Minkowski 
 translations  $\R^4\ni  x=\bl1_2x_0+\rvec\si\rvec x$ 
into time and position  translations is not compatible with the
action of the Lorentz
group $\SO_0(1,3)$. The Dirac distribution has also a nontrivial 
projection for the position   $\SD(2)$ structure
\begin{eq}{l}
2\pi\int dx_0\ep(x_0)
\bl s(x|m)=
m {e ^{- r m}\over r }\cr
\end{eq}and the derived Dirac distribution 
a nontrivial projection for time $\D(1)$ re\-pre\-sen\-ta\-tions
\begin{eq}{l}
\int d^3 x
~i\bl s^{\rm dip}(x| m)=
i{\sin x_0  m-x_0 m\cos x_0 m\over 2 m^2}
\cr
\end{eq}The position projection of the  Dirac distribution leads to 
a Yukawa force which is no  matrix
element of an $\SD(2)$-re\-pre\-sen\-ta\-tion - only of its tangent position
translations $\R^3$.
The time projection of the derived Dirac distribution leads to matrix
elements of reducible nondecomposable $\D(1)$-re\-pre\-sen\-ta\-tions.

Related to the  two Cartan coordinates $\{x_0, r \}$ 
which reflect the rank 2 of the noncompact homogeneous manifold 
$\D(2)$, i.e. two Abelian subgroups $\D(\bl1_2)$ (time) and
$\SO_0(1,1)$ as a dilatation subgroup of the position   manifold $\SD(2)$,
two invariants $\{m_0^2,m_3^2\}$ 
have to characterize the $\D(2)$-representations.
The definite unitary 
 re\-pre\-sen\-ta\-tions $\D(\bl 1_2)\ni e^{x_0\bl1_2}\mape e^{\pm x_0 im_0}
 \in\U(1)$  
are  characterized by a  particle mass $m^2_0$.
A second mass  $m^2_3$ characterizes the indefinite
 unitary re\-pre\-sen\-ta\-tion $\SO_0(1,1)\ni e^{\pm r  }\mape
 e^{\pm r m_3}\in\SU(1,1)$ with an interaction range ${1\over m_3}$
and without particle asymptotics. 
There is no group theoretical reason
to identify both scales $ m^2 _0= m_3^2$ - in general, 
the re\-pre\-sen\-ta\-tions of spacetime $\D(2)$ come with two different
scales whose ratio 
${ m_3^2\over  m_0^2}$ is a re\-pre\-sen\-ta\-tion characteristic 
physically important constant. 
The ratio of the characterizing invariants for particle
and interaction 
should be seen  in analogy to the relative normalization
 of time and position  translations
${\scriptsize\pmatrix{\tau^2&0\cr 0&-\ell^2\bl1_3\cr}}$
as given with the speed of light
$c^2={\ell^2\over\tau^2}$.

\section{Residual Representations} 
 Before the definition of residual representations in general
 their structure will be exemplified  in the familiar example of
 the compact and noncompact abelian groups $\U(1)$ and $\D(1)$.

\subsection{Residual $\U(1)\x\D(1)$-Representations}

An irreducible re\-pre\-sen\-ta\-tion of the complex Abelian group $\exp\C$ 
can be written as residue of its eigenvalue  
by using the 
complex Lie algebra forms $Q\in\C$ 
\begin{eq}{l}
\exp\C\ni   e ^z\mape   e ^{z\ze}=
\oint {dQ\over2i\pi}{1\over Q-\ze}  e ^{zQ},~~\ze\in\irrep\exp \C\cong\C
\end{eq}which, with the canonical conjugation,  gives for  the
irreducible $\U(1)$ and 
$\D(1)$-re\-pre\-sen\-ta\-tions, necessarily in $\U(1)$ 
\begin{eq}{lllll}
\U(1)\ni   e ^{i\al}&\mape   e ^{i\al Z}=&
\int dq \de(q-Z)  e ^{i \al q}
&=\oint {dq\over2i\pi}{1\over q-Z}  e ^{i \al q}&\in\U(1)\cr
&&Z\in\irrep\U(1)&\cong \Z&\cr
\D(1)\ni   e ^{t}&\mape   e ^{t i m}=&
\int dq \de(q-m)  e ^{t i q}&=
\oint {dq\over2i\pi}{ 1\over q- m}  e ^{t iq}&\in\U(1)\cr
&& im\in \irrep\D(1)&\cong i\R&\cr
\end{eq}with the neutral representations  for $Z=0$ and $m=0$ resp. The integrations for the 
compact and noncompact group are related 
to each other via the
Lie algebras and their  forms
by multiplication with the imaginary unit $i$ 
\begin{eq}{c}
\hbox{for compact }\U(1)~~(i\al,q)\lrmap (t,iq)~~\hbox{for noncompact }\D(1)
\end{eq}

Measures of the  integer winding numbers $Z$ as invariants of 
the compact group $\U(1)$
lead to Fourier series  
as measured $\U(1)$-re\-pre\-sen\-ta\-tions
\begin{eq}{rl}
\mu:\irrep \U(1)&\map\R,~~Z\mape\mu(Z)\cr
\meas\irrep\U(1)\ni \mu&\mape D^\mu\in \rep\U(1)\cr
\U(1)\ni   e ^{i\al}&\mape D^\mu(\al)=
{\SUM_{Z\in\Z}}\mu(Z)  e ^{i\al Z}\cr
\end{eq}The continuous 
irreducible re\-pre\-sen\-ta\-tion classes  for $\D(1)$ 
characterized by imaginary numbers $im$ have Lebesque measure $dm$ based
real valued  measures  giving  rise to Fourier integrals 
as measured $\D(1)$-re\-pre\-sen\-ta\-tions
\begin{eq}{rl}
\mu:\irrep \D(1)&\map\R,~~m\mape\mu(m)\cr
\meas\irrep\D(1)\ni\mu&\mape D^\mu\in \rep\D(1)\cr
\D(1)\ni   e ^{t}&\mape
D^\mu(t)=\int dm~\mu(m)  e ^{t i m}\cr
\end{eq}where also matrix elements of reducible nondecomposable re\-pre\-sen\-ta\-tions may
occur by using derivatives with respect to the invariant
\begin{eq}{l}
\mu(m)={\SUM_{N=0,1,\dots}}\mu_N(m)({d\over dm})^N\cr
\end{eq}

\subsection{The Definition of Residual Representations}

Residual representations are complex functions 
on a 
real finite dimensional symmetric space $G$, e.g. a  Lie group,
with tangent space (Lie algebra) $\log G\cong\R^n$,
as above  for $\U(1)$ and  $\D(1)$ and in the following for
$\SU(2)$ and $\SL(\C^2 )$
and  generalized to the
position   manifold  $\SD(2)$ and the spacetime manifold $\D(2)$.

The equivalence classes $\irrep G$ of
the irreducible $G$-re\-pre\-sen\-ta\-tions  
are characterizable by invariants,
taken from a  rational  spectrum  for a compact and  from 
an also continuous spectrum  for
a noncompact Cartan subgroup.
The weights (eigenvalues) for the symmetric space  $G$
are a submodule of the linear forms\footnote{\scriptsize
The linear forms (dual space) of a vector space $V$ are denoted by $V^T$.}
 $q\in (\log G)^T$
of the tangent space $x\in\log G$.
The invariants $\{I_1,\dots ,I_r\}$, characterizing an irreducible re\-pre\-sen\-ta\-tion, 
are related to multilinear tangent space forms
(monomials in the weights).
Appropriate  measures $d^nq~I(q)$  of the linear forms, which 
can be written with a Lebesque measure 
basis and  a distribution of the tangent space forms
$(\log G)^T\cong \R^n$ 
 lead to matrix elements of irreducible 
symmetric space re\-pre\-sen\-ta\-tions 
\begin{eq}{rrlrl} 
&I:\R^n\map&\R,&q \mape& I(q)\cr
&D:\meas\R^n\map & \irrep G,&I\mape& D^I\cr
&D^I:G\map& \C,&g( x)\mape& D^I( x)=\int d^n q I(q)e^{i x q}
\end{eq}The complex generalized functions $I(q)$ have  
poles at the values for the invariants characterizing an 
irreducible representation, the
distributions  come as 
quotients of two polynomials $I(q)={P_N(q)\over P_D(q)}$.
$D^I$ is called a {residual representation of $G$} with
$I(q)$ a {residual group distribution}.

{Measured re\-pre\-sen\-ta\-tions} for a symmetric space (Lie group)  $G$ 
integrate   irreducible $G$-re\-pre\-sen\-ta\-tions
with a measure $d^r I~\mu(I)$ of the invariants
\begin{eq}{rlrl}
\mu:\irrep G\map&\R,&I\mape& \mu(I)\cr
D:\meas\irrep G\map&\rep G,&\mu\mape& D^\mu\cr
D^\mu:G\map& \C,&g( x)\mape& D^\mu( x)=\int d^r I~\mu(I) D^I( x)\cr
\end{eq}The  product in the algebra
of the representation classes  $\rep G$ is implemented via the
convolution
of the distributions for the matrix elements of the
product re\-pre\-sen\-ta\-tion
\begin{eq}{l}
D^{I_1}\ox D^{I_2}=D^{I_1*I_2}
\end{eq}

In the following, these general structures will be concretized 
for the groups and symmetric spaces relevant for the spacetime model $\D(2)$.

\subsection
{Residual $\SO(2)\x\SO_0(1,1)$-Representations}

The  real Abelian
group $\SO(2)\x\SO_0(1,1)$ has its  irreducible  selfdual complex 
re\-pre\-sen\-ta\-tions in the 
two types of 2-di\-men\-sio\-nal unitary groups,
the definite unitary $\SU(2)$ 
or the indefinite unitary  $\SU(1,1)$
\begin{eq}{rcl}
 \SO(2)\x\SO_0(1,1)&\map&
\left\{
\begin{array}{ll}
\SO(2)&\subnoteq\SU(2)\cr
\SO_0(1,1)&\subnoteq\SU(1,1)\cr\end{array}\right.
\cr
  e ^{( i\al+x)\si^3}
&\mape&   e ^{(i\al Z+ x\de)\si^3}\cr
\end{eq}The unitary  groups $\SU(2)$ and $\SU(1,1)$  define the 
weights $(Z,\de)$ of the principal (compact)
 and supplementary (noncompact) 
re\-pre\-sen\-ta\-tions resp.

The principal  $\SO(2)\x\SO_0(1,1)$-weights coincide with the 
$\U(1)\x\D(1)$-weights $\Z\x i\R$.
An integer eigenvalue pair
$\{\pm Z\}$
characterizes a selfdual $\SO(2)$-re\-pre\-sen\-ta\-tion 
 \begin{eq}{l}
 \SO(2)\ni{\scriptsize\pmatrix{
 \cos \al& i\sin \al\cr 
 i\sin \al&\cos\al\cr}}
 \mape 
{\scriptsize\pmatrix{
 \cos \al Z&i\sin\al Z\cr 
 i\sin \al Z&\cos \al Z\cr}}\cong
{\scriptsize\pmatrix{
 e^{i\al Z}&0\cr 
 0&e^{-i\al Z}\cr}}\in \SU(2)
\end{eq}leading to a quadratic natural number valued invariant 
$Z^2$. An
imaginary  continuous eigenvalue pair
 $\{\pm im\}$ characterizes
 a selfdual compact $\SO_0(1,1)$-re\-pre\-sen\-ta\-tion 
 \begin{eq}{l}
 \SO_0(1,1)\ni{\scriptsize\pmatrix{
 \cosh x&\sinh x\cr 
 \sinh x&\cosh x\cr}}
\mape 
{\scriptsize\pmatrix{
 \cos xm&i\sin xm\cr 
 i\sin xm&\cos xm\cr}}\cong
{\scriptsize\pmatrix{
 e^{xim}&0\cr 
 0&e^{-xim}\cr}}\in\SU(2)
\end{eq}with  a continuous positive  invariant $m^2\ge 0$
\begin{eq}{rllrll}
\weights\SO(2)\hskip-2mm&=\{Z\}&\hskip-2mm\cong\Z,&
\irrep\SO(2)\hskip-2mm&=\{|Z|\}&\hskip-2mm\cong\N_0\cr
\weights^{(2,0)}\SO_0(1,1)\hskip-2mm&=\{im\}&\hskip-2mm\cong i\R,&
\irrep^{(2,0)}\SO_0(1,1)\hskip-2mm&=\{m^2\}&\hskip-2mm\cong\R^+\cr
\end{eq}

The new real $\SO_0(1,1)$-weights $m\in\R$ (supplementary)
in contrast to the imaginary  principal weights $im\in i\R$ 
above come for dimensions
$n\ge2$ with the possibility of  
indefinite unitary groups. 
A supplementary $\SO_0(1,1)$-re\-pre\-sen\-ta\-tion  
is characterized by a real continuous eigenvalue pair $\{\pm m\}$
 \begin{eq}{l}
 \SO_0(1,1)\ni{\scriptsize\pmatrix{
 \cosh x&\sinh x\cr 
 \sinh x&\cosh x\cr}}
 \mape 
{\scriptsize\pmatrix{
 \cosh xm&\sinh xm\cr 
 \sinh xm&\cosh xm\cr}}
\cong {\scriptsize\pmatrix{
 e^{xm}&0\cr 
 0&e^{-xm}\cr}}\in \SU(1,1)
 \end{eq}with a continuous  negative definite invariant
\begin{eq}{l}
\weights^{(1,1)}\SO_0(1,1)\hskip-1mm=\{m\}=\R,~~
\irrep^{(1,1)}\SO_0(1,1)\hskip-1mm=\{-m^2\}\cong\R^-\cr
 \end{eq}

Residual re\-pre\-sen\-ta\-tions in $\SO(2)$ (principal)
with invariants $m^2\in\R^+$  can be formulated by distributions with 
the $q$-integration deformed as prescribed by  $q^2\mp io$ 
which for an undeformed integration 
gives  singularities at  $m^2\pm io=(|m|\pm io)^2$ 
\begin{eq}{l}
\begin{array}{rll}
 e ^{\pm i|t m|}&= \int d^1q~ [m^2]^0_\pm(q)e^{it q},&
[m^2]^0_\pm(q)=\pm{1\over i\pi} { |m|\over q^2 \mp io-m^2}\cr
 \ep(t) e ^{\pm i|t m|}&= \int d^1q~ [m^2]^1_\pm(q)e^{it q},&
[m^2]^1_\pm(q)={1\over i\pi} { q\over q^2 \mp io-m^2}\end{array}\cr
\hbox{for }\SO(2),\SO_0(1,1)\map\SU(2),~~m\in(\Z,\R)\cr
\end{eq}

Residual re\-pre\-sen\-ta\-tions in $\SU(1,1)$ (supplementary)
with invariants $-m^2\in\R^-$ are obtained  
from residual re\-pre\-sen\-ta\-tions in $\SU(2)$ (principal)
by the real-imaginary exchange $(it,q)\lrmap(x,iq)$
\begin{eq}{l}
\begin{array}{rll}
e ^{-|xm|}&=\int d^1q~[-m^2]^0(q)  e ^{-xiq},&
[-m^2]^0(q) ={1\over\pi}{|m|\over q^2+m^2} \cr
-\ep(x)e ^{-|xm|}&=\int d^1q~[-m^2]^1(q)  e ^{-xiq},&
[-m^2]^1(q) ={1\over i\pi}{q\over q^2+m^2}\end{array} \cr
\hbox{for }\SO_0(1,1)\map\SU(1,1),~~m\in\R\cr
\end{eq}

In the transition from the compact to the noncompact re\-pre\-sen\-ta\-tion
structure the invariant $\pm i|m|$ has to be replaced by $-|m|$
\begin{eq}{l}
\hbox{for }\SO(2)\subnoteq\SU(2)~~\pm i|m|\lrmap -|m|\hbox{ for
}\SO_0(1,1)\subnoteq\SD(2)
\end{eq}

The matrix elements for the re\-pre\-sen\-ta\-tions in $\SO(2)$ and $\SO_0(1,1)$
fulfill the 2nd order differential equations
\begin{eq}{l}
({d^2\over d t^2}+m^2)e^{\pm i| t m|}=\pm 2i|m|\de(t),~~
({d^2\over d x^2}-m^2)e^{-|xm|}=-2|m|\de(x)\cr
\end{eq}

The product re\-pre\-sen\-ta\-tions arise by  convolution 
- for $\SO(2)$ with equal type, either  $+ io$ or $-io$ -
with the supindices $\{1,0\}$ adding up modulo 2, e.g.
\begin{eq}{l}
\left.\begin{array}{cccl}
[m_1^2]^1_\pm&*& ~[m_2^2]^1_\pm&=~[m_+^2]^0_\pm\cr
[-m_1^2]^1&*& [-m_2^2]^1&=[-m_+^2]^0\cr\end{array}\right\}~~ |m_+|=|m_1|+|m_2|\cr
\end{eq}With the convolution the distributions
\begin{eq}{rl}
\irrep\SO(2)&=\{q\mape [Z^2]^1_\pm(q)
={1\over i\pi}{ q\over q^2\mp io-Z^2}
\mid Z\in\Z\}\cr
\irrep^{(2,0)}\SO_0(1,1)&=\{q\mape [m^2]^1_\pm(q)
={1\over i\pi}{ q\over q^2\mp io-m^2}
\mid m\in\R\}\cr
\irrep^{(1,1)}\SO_0(1,1)&=\{q\mape [-m^2]^1(q)
={1\over i\pi}{ q\over q^2+m^2}
\mid m\in\R\}\cr
\end{eq}generate the compact and noncompact selfdual Abelian
re\-pre\-sen\-ta\-tions resp. The neutral representations arise for trivial
invariant.

\subsection{Residual Representations for Spin $\SU(2)$}

If the compact group 
$\SO(2)$ comes as Cartan subgroup in the special group 
$e^{-i\rvec x\rvec\si}\in \SU(2)$
 with the Cartan polar decomposition
\begin{eq}{l}
\SU(2)\cong\SO(2)\x\SO(3)/\SO(2)\cr
\end{eq}residual re\-pre\-sen\-ta\-tions
employ the  forms $\rvec q\in\R^3$ of the 
tangent Lie algebra $\log\SU(2)$ (angular momenta) 
with the singularities of the distributions
determined by the values of the invariant 
bilinear Killing form $\rvec q^2$
as singularity location of a  {dipole}  
\begin{eq}{rl}
\hbox{for }\SO(2)\subnoteq\SU(2):
~~ e ^{\pm i r| m|}&=\int d^3q~ [0,m^2]_\pm(\rvec q)
 e ^{-i\rvec x\rvec q}\cr
[0,m^2]_\pm(\rvec q)&=\pm{1\over i\pi^2}{ |m|\over (\rvec q^2\mp io-m^2)^2} 
,~~m\in\R\cr
\end{eq}This scalar re\-pre\-sen\-ta\-tion and similar  integrals  can be obtained by
derivations with respect 
to the invariant $m^2$  and the Lie parameter $\rvec x$
from the in- and outgoing {spherical waves} 
\begin{eq}{rl}
\int {d^3q\over2 \pi^2}{1\over \rvec q^2\mp io-m^2}  e ^{-i\rvec x\rvec q}
={  e ^{ \pm i r| m|}\over r },&m\in\R,~\rvec x\ne0\cr
{\p\over \p m^2}={1\over2|m|}{\p\over\p |m|},~~
{\p\over\p\rvec x}={\rvec x\over r }{\p\over\p r },&~~
({\p^2\over \p \rvec x^2}+m^2)
 {e^{\pm i r|  m|}\over r }=4\pi \de(\rvec x)\cr
\end{eq}which, however, are  no 
$\SU(2)$-re\-pre\-sen\-ta\-tion
matrix elements  because of the Lie parameter $\rvec x=0$
singularity.

The 
scalar matrix elements
fulfill 4th order differential equations
\begin{eq}{l}
~~({\p ^2\over \p\rvec x^2}+m^2)^2e^{\pm i r| m|}
=\mp 8\pi i|m|\de(\rvec x)\cr
\end{eq}

Vector valued   distributions represent nontrivially  the 2-sphere 
$\SO(3)/\SO(2)$
\begin{eq}{l}
-{\rvec x\over r }  e ^{\pm i r| m|}=
\int d^3q~ [1,m^2]_\pm(\rvec q)
  e ^{-i\rvec x\rvec q}=\int{d^3q\over i\pi^2}
  {\rvec q\over (\rvec q^2\mp io-m^2)^2}
  e ^{-i\rvec x\rvec q},~~m\in\R\cr
\end{eq}leading to the matrix elements of the defining Pauli 
re\-pre\-sen\-ta\-tion 
\begin{eq}{c}
\left.\begin{array}{ll}
\int d^3 q~[0,1]_\pm(\rvec q)e^{-i\rvec x\rvec q}&=e ^{\pm i r }\cr
\int d^3 q~[1,1]_\pm(\rvec q)e^{-i\rvec x\rvec q}&
= -{\rvec x\over r }  e ^{\pm i r }\end{array}\right\}
\llrmap
 e ^{-i\rvec x\rvec \si}
=\bl1_2\cos r 
-{\rvec\si\rvec x\over r }i\sin r 
\end{eq}The spherical dependence ${\rvec x\over r }$
replaces the $\ep(x)$-dependence for $\SO(2)$.

With the Lie algebra additive 
convolution  product      
of the distributions for the
{irreducible residual $\SU(2)$-re\-pre\-sen\-ta\-tions}
\begin{eq}{l}
\irrep\SU(2)=\{\rvec q\mape [1,m^2]_\pm(\rvec q)
={1\over i\pi^2}{\rvec q\over (\rvec q^2\mp io-m^2)^2}
\mid |m|=2J\in\N_0\}\cr
\end{eq}involving the neutral representation for trivial invariant $m=0$
one can combine the matrix elements for all other re\-pre\-sen\-ta\-tions,
e.g. the scalar ones with $|m_1|+|m_2|=|m_+|$
\begin{eq}{rl}
{  x^a\over r }  e ^{\pm i r| m_1|}\de_{ab}
{ x^b\over r }  e ^{\pm i r| m_2|}
&= e ^{\pm i r| m_+|}\cr
[1,m_1^2]_\pm\stackrel{J'=0}*[1,m_2^2]_\pm(\rvec q)&
=[0,m_+^2]_\pm(\rvec q)\cr
=({1\over i\pi^2})^2\int d^3q_1 d^3q_2{q_1^a\over (\rvec q_1^2\mp io-m_1^2)^2}
\de(\rvec q_1+\rvec q_2-\rvec q)\de_{ab}{q_2^a\over 
(\rvec q_2^2\mp io-m_2^2)^2}
&=\pm{1\over i\pi^2}
{|m_+|\over (\rvec q^2\mp io-m_+^2)^2}\cr
\end{eq}or for
the  adjoint re\-pre\-sen\-ta\-tion 
\begin{eq}{l}
\de_{ab}\cos2 r +{ x_a x_b\over r^2}(1-
\cos2 r )+\ep_{abc}{ x_c\over r }\sin 2 r 
\end{eq}which arises
for $|m_+|=|m_1|+|m_2|=2$
\begin{eq}{ll}
{  x^a\over r }  e ^{\pm i r| m_1|}
{ x^b\over r }  e ^{\pm i r| m_2|}
&={ x^a x^b\over r ^2}  e ^{\pm i r| m_+|}\cr
[1,m_1^2]_\pm\stackrel{2J'=2}*[1,m_2^2]_\pm(\rvec q)
&=({1\over i\pi^2})^2\int d^3q_1 d^3q_2{q_1^a\over (\rvec q_1^2\mp io-m_1^2)^2}
\de(\rvec q_1+\rvec q_2-\rvec q){q_2^a\over (\rvec q_2^2\mp io-m_2^2)^2}
\cr
\end{eq}

In general, the matrix elements of $\SU(2)$-re\-pre\-sen\-ta\-tions come as
products of a homogeneous polynomial (spherical harmonics)
of degree $2J'$ for the sphere $\SO(3)/\SO(2)$-re\-pre\-sen\-ta\-tion
and an expontential for the Cartan subgroup $\SO(2)$
with winding numbers $\pm 2J$
\begin{eq}{c}
\{[{\rvec x\over  r }]^{2J'}e^{\pm i r 2J}\mid
2J'\in\N_0,~2J\in\N_0\}\cr
[{\rvec x\over  r }]^0=\{1\},~~
[{\rvec x\over  r }]^1=\{{ x^a\over  r }\mid a=1,2,3\},
~~[{\rvec x\over  r }]^2=
\{{ x^a x^b\over r^2}-{\de^{ab}\over3}\},\dots\cr
\end{eq}
 
Matrix elements of  measured $\SU(2)$-re\-pre\-sen\-ta\-tions 
use  real measures of the
irreducible re\-pre\-sen\-ta\-tions classes
\begin{eq}{rl}
\mu:\irrep\SU(2)&\map  \R,~~2J\mape \mu(4J^2)\cr
\meas\irrep\SU(2)\ni\mu&\mape  D_\pm^\mu\in \rep\SU(2)\cr
\end{eq}with the functions on the spin group  $\SU(2)$
\begin{eq}{rl}
\SU(2)\ni e^{i\rvec x\rvec\si}\mape D_\pm^\mu(\rvec x)&
={\SUM_{2J=0,1,\dots}}\mu(4J^2)\int {d^3 q\over i\pi^2}
{\rvec q\over (\rvec q^2\mp io-4J^2)^2}
e^{-i\rvec x\rvec q}\cr
&=-{\rvec x\over  r } {\SUM_{2J=0,1,\dots}}\mu(4J^2) 
e^{\pm i r 2J}
\end{eq}

\subsection{Residual Representations for Position $\SD(2)$}

For the position   manifold $e^{-\rvec x\rvec \si}\in\SD(2)$ 
with the Cartan polar decomposition
\begin{eq}{l}
\SD(2)\cong\SO_0(1,1)\x\SO(3)/\SO(2)\cr
\end{eq}residual 
re\-pre\-sen\-ta\-tions use the 
tangent space forms  (momenta $\rvec q\in\R^3$)
and, in comparison  to $\SU(2)$, the tangent space 
real-imaginary exchange  for compact-noncompact
\begin{eq}{l}
\hbox{for }\SU(2)~~\left\{\begin{array}{rl}
\hbox{Lie algebra and forms }(i\rvec x,\rvec q)
&\lrmap 
(\rvec x,i\rvec q)\cr
\hbox{invariant }\pm i|m|&\lrmap -|m|\cr\end{array}\right\}
~~\hbox{for }\SD(2)
\end{eq}

As for the Cartan subgroup $\SO_0(1,1)$ there exist two types:
The compact re\-pre\-sen\-ta\-tions $\SD(2)\map \SU(2)$ (principal)
with $\SO_0(1,1)\map \SO(2)$
and the noncompact ones $\SD(2)\map \SU(1,1)$ (supplementary)
with  faithful re\-pre\-sen\-ta\-tions $\SO_0(1,1)\map \SO_0(1,1)$. 
Both are re\-pre\-sen\-ta\-tions of the homogeneous position manifold
 in a unitary group,
definite or indefinite.

From the {Yukawa potential} 
\begin{eq}{l}
\int {d^3 q\over 2\pi^2}{1\over \rvec q^2+m^2}  e ^{-\rvec x i\rvec q}
={  e ^{- r| m|}\over  r }
,~~m\in\R,~\rvec x\ne0\cr
({\p^2\over \p \rvec x^2}-m^2) {e^{- r| m|}\over r }
=-4\pi \de(\rvec x)\cr
\end{eq}which, by itself, is no $\SD(2)$-re\-pre\-sen\-ta\-tion
matrix element 
because of the $\rvec x=0$ singularity, one obtains
by derivations ${\p\over\p m^2}$ and ${\p\over\p\rvec x}$ 
the scalar matrix elements, trivially representing the sphere $\SO(3)/\SO(2)$
\begin{eq}{rl}
\hbox{for }\SO_0(1,1)\subnoteq\SD(2):
~~e ^{- r| m|}&=
\int d^3 q~[0,-m^2](\rvec q) e ^{-\rvec x i\rvec q}\cr
[0,-m^2](\rvec q)&={1\over\pi^2}{|m|\over (\rvec q^2+m^2)^2}, ~~m\in\R
\cr
\end{eq}and the  {fundamental noncompact
residual $\SD(2)$-re\-pre\-sen\-ta\-tions} 
using a vector valued distribution
\begin{eq}{l}
-{\rvec x\over r }  e ^{- r| m|}=
\int d^3 q~ [1,-m^2](\rvec q) e ^{-\rvec x i\rvec q}
=\int {d^3 q\over i\pi^2}{\rvec q\over (\rvec q^2+m^2)^2}
e ^{-\rvec x i\rvec q} ,~~m\in\R\cr
\end{eq}This has to be compared with the elements in
the defining  re\-pre\-sen\-ta\-tion
\begin{eq}{l}
  e ^{-\rvec x \rvec \si}
=\bl1_2\cosh r 
-{\rvec\si\rvec x\over r }\sinh r 
\end{eq}

The  scalar matrix elements
fulfill 4th order differential equations
\begin{eq}{rl}
({\p^2\over \p \rvec x^2}-m^2)^2 e^{- r| m|}
&= 8\pi |m|\de(\rvec x)\cr
({\p^2\over \p \rvec x^2}+m^2)^2 e^{\pm i r| m|}
&= \mp 8\pi i |m|\de(\rvec x),~~m\in\R\cr
\end{eq}

In contrast to the spin group $\SU(2)$ where the re\-pre\-sen\-ta\-tions
of the compact  Cartan subgroup $\SO(2)$ 
and the sphere $\SO(3)/\SO(2)$ go both with   
discrete invariants $2J',2J\in\N_0$ -  
arising  as  degree 
of the spherical harmonics and as winding numbers,
the continuous invariant $m^2\in\R^+$ of the noncompact
Cartan group $\SO_0(1,1)$-re\-pre\-sen\-ta\-tion 
in the case of the position   manifold
$\SD(2)$ 
is taken from a different spectrum as the discrete invariant 
$2J'\in\N_0$ for the sphere 
$\SO(3)/\SO(2)$-re\-pre\-sen\-ta\-tions.
Again
the convolution products of the distributions for
the {fundamental residual $\SD(2)$-re\-pre\-sen\-ta\-tions}
\begin{eq}{rl}
\irrep^{(1,1)}\SD(2)&=\{\rvec q\mape[1,-m^2](\rvec q)={1\over i\pi^2}
{\rvec q\over (\rvec q^2+m^2)^2}\mid m\in\R\}\cr
\irrep^{(2,0)}\SD(2)&=\{\rvec q\mape[1,m^2]_\pm (\rvec q)={1\over i\pi^2}
{\rvec q\over (\rvec q^2\mp io -m^2)^2}\mid m\in\R\}\cr
\end{eq}define the matrix elements of the $\SD(2)$-re\-pre\-sen\-ta\-tions.
The representations for trivial invariant $m=0$ will be called neutral.

Measured $\SD(2)$-re\-pre\-sen\-ta\-tions 
use real measures of the 
continuous invariants
\begin{eq}{rl}
\mu:\irrep\SD(2)&\map \R,~~m^2\mape\mu(m^2)\cr
\meas\irrep\SD(2)&\ni\mu\mape D^\mu\in \rep\SD(2)\cr
\end{eq}with the functions on the position   manifold $\SD(2)$
\begin{eq}{l}
\SD(2)\ni e^{-\rvec x\rvec\si}\mape 
D^\mu(\rvec x)=\left\{\begin{array}{l}
\int_0^\infty  dm^2\mu(m^2) 
\int {d^3 q\over i\pi^2}{\rvec q \over (\rvec q^2+m^2)^2}
  e ^{-\rvec x i\rvec q}\cr
 = -{\rvec x\over r }\int_0^\infty  dm^2\mu(m^2) 
 e ^{- r| m|}\hbox{ for }\rep^{(1,1)}\SD(2)\cr
 \hbox{ and } \cr
\int_0^\infty  dm^2\mu(m^2) 
\int {d^3 q\over i\pi^2}{\rvec q \over (\rvec q^2\mp io -m^2)^2}
  e ^{-\rvec x i\rvec q}\cr
=-  {\rvec x\over r }\int_0^\infty  dm^2\mu(m^2) 
 e ^{\pm i r| m|}\hbox{ for }\rep^{(2,0)}\SD(2)
\end{array}\right.  \cr
\end{eq}

The two integrations 
in measured re\-pre\-sen\-ta\-tion matrix elements  go over the tangent space forms 
$\int d^3q$
and the invariants $\int_0^\infty dm^2$
with the dimensions $3$ and $1$ of 
the symmetric space $\SD(2)$  and a Cartan subgroup $\SO_0(1,1)$ resp. 
Matrix elements of reducible nondecomposable re\-pre\-sen\-ta\-tions 
occur by using derivatives with respect to the invariant
\begin{eq}{l}
\mu(m^2)={\SUM_{N=0,1,\dots}}\mu_N(m^2)({d\over dm^2})^N\cr
\end{eq}

\section{Residual Representations of Spacetime}

Matrix elements of re\-pre\-sen\-ta\-tions of a symmetric space (Lie group) 
can be formulated  as residues for characterizing invariant singularities
of their tangent translation (Lie algebra) forms.
For the groups $\U(1)$, $\D(1)$, $\SU(2)$  and the position   manifold   
$\SD(2)$, as done in the former sections, this is only a
reformulation of  known structures.
Residual re\-pre\-sen\-ta\-tions constitute a genuine formulation for 
 the  rank 2
symmetric spacetime
$\D(2)$. 
Two values for the Lorentz invariant energy-momentum square $q^2$ characterize 
the action of the causal group $\D(1)$ and the position manifold $\SD(2)$.

Representations of spacetime 
\begin{eq}{rl}
\D(2)&=\D(\bl 1_2)\x\SD(2)
\cong\D(\bl1_2)\x\SO_0(1,3)/\SO(3)\cr
&\cong \D(\bl1_2)\x\SO_0(1,1)\x \SO(3)/\SO(2)\cr
\end{eq}will be formulated as Fourier transforms of
energy-momentum distributions,
compatible with the action of the Lorentz group 
$\SO_0(1,3)$ on the tangent Minkowski spacetime.  
The two invariant masses  
characterizing the re\-pre\-sen\-ta\-tions 
are implemented via singularities.

The {irreducible residual re\-pre\-sen\-ta\-tion 
matrix elements of spacetime $\D(2)$},
parametrizable  with causal vectors  $x\vth(x^2)\in\R^4$ in 
tangent Minkowski spacetime
where the two reflected points $\{\pm x\vth(x^2)\}$ 
and, equally,  their re\-pre\-sen\-ta\-tion images have to be identified 
\begin{eq}{l}
\D(2)\ni x\vth(x^2) \mape (m_0^2;1,-m_3^2) (x)
=\int d^4 q ~[m_0^2;1,-m_3^2](q)e^{xiq} \cr
\end{eq}involve a two factorial  energy-momentum distribution
\begin{eq}{l}
\irrep\D(2)=\{q\mape [m_0^2;1,-m_3^2] (q)
={1\over i\pi^3}{2q \over   (q_\ro P^2-m_0^2)(q_\ro P^2-m_3^2)^2}
\mid m_0,m_3\in\R\}
\end{eq}It
describes the Lorentz compatible embedding 
for the re\-pre\-sen\-ta\-tion of the 
two $\D(2)$-factors and  involves a simple pole   (particle singularity)
for  the  compact  re\-pre\-sen\-ta\-tion of
a Cartan subgroup time 
\begin{eq}{rl}
\hbox{for }\D(\bl1_2)\map \SU(2):&
{1 \over   q_\ro P^2-m_0^2}\cr
\hbox{with pole location for $\rvec q=0$:}&
q_0^2=m_0^2\cr
\end{eq}and 
 a dipole (interaction singularity)
 for the noncompact  re\-pre\-sen\-ta\-tion of
the position   symmetric space $\SD(2)$ with  Cartan subgroup  $\SO_0(1,1)$
\begin{eq}{rl}
\hbox{for }\SD(2)\supnoteq\SO_0(1,1)\map \SU(1,1):&
{1\over (q_\ro P^2-m_3^2)^2}\cr
\hbox{with dipole location for $q_0=0$:}&\rvec q^2=-m_3^2\cr
\end{eq}The 2-sphere 
$\SO(3)/\SO(2)$ nontrivially representing factor
is given by $\{q^j\}_{j=0}^3$ in the numerator.

The  Fourier transform of a principal  value distribution
is  the causal Fourier transform of a  Dirac distribution
and vice versa
\begin{eq}{l} \int {d^4 q\over i\pi}{1\over q_\ro P^2-m^2}
{\scriptsize\pmatrix{
 e^{xiq}\cr
\ep(x_0)e^{xiq}\cr}}=\int d^4 q~ \ep(q_0)\de(q^2-m^2){\scriptsize\pmatrix{
\ep(x_0)e^{xiq}\cr
 e^{xiq}\cr
}}
\end{eq}

The matrix elements of the  {measured spacetime re\-pre\-sen\-ta\-tions}
as $\D(2)$-functions 
involve a  measure for the two continuous
invariants  
\begin{eq}{l}
\mu:\irrep\D(2)\map\R,~~(m_0^2,m_3^2)\mape
\mu(m^2_0,m^2_3)\cr
\meas\irrep\D(2)\ni\mu\mape D^\mu\in\rep\D(2)\cr
\D(2)\ni x\vth(x^2)\mape D^\mu(x)=\int_0^\infty d m_0^2d m_3^2~
\mu(m_0^2,m_3^2)
\int {d^4q\over i\pi^3} {2q\over 
(q_\ro P^2-m_0^2)(q_\ro P^2-m_3^2)^2}  e ^{xiq}
\end{eq}

The $\D(2)$-re\-pre\-sen\-ta\-tions are different from
the Lorentz compatible position   distributions of time re\-pre\-sen\-ta\-tions
as used for the quantization of the tangent Minkowski spacetime 
particle fields (K\"allen-Lehmann re\-pre\-sen\-ta\-tions\cite{KL}), e.g. for 
a spin ${1\over2}$ massive particle in a Dirac field
\begin{eq}{l}
\hbox{particle fields: }
\int_0^\infty d m^2
\mu(m^2)\int {d^4q\over \pi^3} {i(\ga q+|m|)\over q^2-m^2+io}  e ^{xiq},~~
\mu(m^2)\ge0
\end{eq}with probability related 
spectral measure $\mu(m^2)$ for the invariant of the definite
unitary re\-pre\-sen\-ta\-tions of the spacetime translations.

One obtains from the Lorentz scalar spacetime distribution with a simple 
energy-momentum pole
\begin{eq}{l}
(\p^2+m^2)\int {d^4q\over\pi^3}{1\over q^2_\ro P-m^2 }e^{xiq}
=-16\pi\de(x)
\end{eq}the derivatives ${\p\over \p m^2}$  with respect to  the invariant 

\begin{eq}{lll}
\int {d^4q\over \pi^3}{\Ga(2+N)\over (q^2_\ro P-m^2)^{2+N} }e^{xiq}
&=\left\{
\begin{array}{lll}
{\p\over\p{x^2\over 4}}&\vth(x^2)\cl E_0({x^2m^2\over4}),&N=-1\cr
\({\p\over\p m^2}\)^N&\vth(x^2)\cl E_0({x^2m^2\over4}),&N=0,1,\dots\cr
\end{array}\right.\cr
\cr
&=\left\{
\begin{array}{ll}
-\de({x^2\over4})+\vth(x^2)m^2\cl E_1({x^2m^2\over4}),&N=-1\cr
\vth(x^2)(-{x^2\over4})^N\cl E_N({x^2m^2\over4}),&N=0,1,\dots\cr
\end{array}\right.\cr
\end{eq}and the derivative  ${\p\over \p x}$ with respect to the Lie parameter 
\begin{eq}{lll}
\int {d^4q\over i\pi^3}{q~\Ga(3+N)\over (q^2_\ro P-m^2)^{3+N} }e^{xiq}
&={x\over2}\left\{
\begin{array}{lll}
\({\p\over\p{x^2\over 4}}\)^N&\vth(x^2)\cl E_0({x^2m^2\over4}),&N=-1,-2\cr
\({\p\over\p m^2}\)^N&\vth(x^2)\cl E_0({x^2m^2\over4}),&N=0,1,\dots\cr
\end{array}\right.\cr
\cr
&={x\over2}\left\{\begin{array}{ll}
\de'({x^2\over4})-m^2\de({x^2\over4})+
\vth(x^2)m^4\cl E_2({x^2m^2\over4}),&N=-2\cr
-\de({x^2\over4}) +\vth(x^2)m^2\cl E_1({x^2m^2\over4}),&N=-1\cr
\vth(x^2)(-{x^2\over4})^N\cl E_N({x^2m^2\over4}),&N=0,1,\dots\cr
\end{array} \right.
\end{eq}which involve the Bessel functions $\cl J_N$ with 
$\xi\in\R$
\begin{eq}{rcrcrl}
\cl E_N({\xi^2\over 4})&=&{\cl J_N(\xi)\over({\xi\over2})^N}
&=&(-{\p\over\p{\xi^2\over 4}})^N\cl J_0(\xi)
&={\SUM_{n=0}^\infty}{(-{\xi^2\over 4})^n\over n!(N+n)!}\cr
(-{\xi^2\over 4})^N\cl E_N({\xi^2\over 4})&=&(-{\xi\over2})^N\cl J_N(\xi)
&=&({\xi^2\over 4})^N({\p\over\p{\xi^2\over 4}})^N\cl J_0(\xi)&\cr
\end{eq}

The distributions for
strictly negative nildimension $N$, i.e. 
with a  Dirac distribution on the light cone $x^2=0$, 
are no spacetime $\D(2)$-re\-pre\-sen\-ta\-tion
matrix elements. One  obtains for the irreducible spacetime re\-pre\-sen\-ta\-tions
the $\D(2)$-functions
\begin{eq}{rl}
x\vth(x^2)\mape (m_0^2;1,-m_3^2) (x)&=
\int {d^4 q\over i\pi^3}
{2q \over   (q_\ro P^2-m_0^2)(q_\ro P^2-m_3^2)^2}e^{xiq}\cr
&=x\vth(x^2) \Brack{
{
m_0^4\cl E_2({x^2m_0^2\over4})
-m_3^4\cl E_2({x^2m_3^2\over4})
\over (m_0^2-m_3^2)^2}+
{m_3^2\cl E_1({x^2m_3^2\over4})\over m_3^2-m_0^2} }\cr
\end{eq}The neutral elements, either for $\SD(2)$ or $\D(1)$, are defined by
trivial masses
\begin{eq}{rll}
(m_0^2;1,0) (x)&=
\int {d^4 q\over i\pi^3}
{2q \over   (q_\ro P^2-m_0^2)(q_\ro P^2)^2}e^{xiq}
&=x\vth(x^2)\cl E_2({x^2m_0^2\over4})\cr
 (0;1,-m_3^2) (x)&=
\int {d^4 q\over i\pi^3}
{2q \over   q_\ro P^2(q_\ro P^2-m_3^2)^2}e^{xiq}
&=x\vth(x^2) [-\cl E_2({x^2m_3^2\over4})
+\cl E_1({x^2m_3^2\over4})]\cr
(0;1,0) (x)&=
\int {d^4 q\over i\pi^3}
{2q \over   (q_\ro P^2)^3}e^{xiq}&={x\over2}\vth(x^2)\cr
\end{eq}

\section{Associated Residual Distributions}

Given a   residual distribution $I(q)=I_0(q)$
for 
an irreducible   re\-pre\-sen\-ta\-tion of a symmetric space (Lie group) $G$, 
singular distributions $\{I_N(q)\}$
using the same  pole locations, but with possibly different orders,
are called {$I$-associated residual distributions}. 
The possibly different singularity orders of the associated 
distributions 
 will be characterized by  integer  {nildimensions} $N\in\Z$. 

Associated to the Dirac distribution for an irreducible Abelian group re\-pre\-sen\-ta\-tion
are its derivatives
\begin{eq}{l}
\{\de^{(N)}(m-q)
\mid N=0,1,\dots\}
\left\{\begin{array}{ll}
m\in\Z&\hbox{for }\irrep\U(1)\cr
m\in\R&\hbox{for }\irrep\D(1)\end{array}\right.\cr
\int dq~\de^{(N)}(m-q)e^{tiq}=
\oint {dq\over2i\pi}{\Ga(1+N)\over (q-m)^{1+N}}e^{tiq}=
(i t)^Ne^{tim}\cr
\end{eq}For the selfdual Abelian groups one has as associatiated distributions
for the compact representations (always only where the $\Ga$-functions are defined)
\begin{eq}{rl}
\{{1\over i\pi}{q~\Ga(1+N)\over (q^2\mp io-m^2)^{1+N}}\}
\hbox{ with}
&\left\{\begin{array}{ll}
m\in\Z&\hbox{for }\irrep\SO(2)\cr
m\in\R&\hbox{for }\irrep^{(2,0)}\SO_0(1,1)\end{array}\right.\cr

\int{d q\over i\pi}{q~\Ga(1+N)\over
(q^2\mp io-m^2)^{1+N}}e^{-i x q}&=-\ep(x)
({\p\over\p m^2})^{N}e^{\pm i| x m|}\cr
&=\left\{\begin{array}{l}
-\ep(x)e^{\pm i| x m|},~\mp i{ x\over 2|m|}e^{\pm i| x m|}
,\dots\cr
\hbox{for }N=0,1,\dots \cr\end{array}\right.\cr

\pm\int{d q\over i\pi}{|m|~\Ga(1+N)\over
(q^2\mp io-m^2)^{1+N}}e^{-i x q}&=|m|({\p\over\p m^2})^N
{e^{\pm i| x m|}\over|m|}\cr
&=\left\{\begin{array}{l}
e^{\pm i| x m|}
,~-{1\mp i|  x m|\over 2m^2}e^{\pm i| x m|},\dots\cr
\hbox{for }N=0,1,\dots \cr\end{array}\right.\cr
\end{eq}and for the noncompact $\SO_0(1,1)$-representations
\begin{eq}{rl}
 \{{1\over i\pi}{q~\Ga(1+N)\over (q^2+m^2)^{1+N}}\}
 \hbox{ with}
&m\in\R
\hbox{ for }\irrep^{(1,1)}\SO_0(1,1)\cr
\int{d q\over i\pi}{ q~\Ga(1+N)\over
(q^2+m^2)^{1+N}}e^{-i x q}&=-\ep(x)
(-{\p\over\p m^2})^{N}e^{-| x m|}\cr
&=\left\{\begin{array}{l}
-\ep(x)e^{-| x m|},~-{  x\over 2|m|}e^{-| x m|}
,\dots\cr
\hbox{for }N=0,1,\dots \cr\end{array}\right.\cr
\int{d q\over \pi}{|m|~\Ga(1+N)\over
(q^2+m^2)^{1+N}}e^{-i x q}&=|m|(-{\p\over\p m^2})^{N}{e^{-| x m|}\over|m|}\cr
&=\left\{\begin{array}{l}
e^{-| x m|},~{1+|  x m|\over 2m^2}e^{-| x m|}
,\dots\cr
\hbox{for }N=0,1,\dots \cr\end{array}\right.\cr

\end{eq}

With respect to the sign of the {nildimension} $N$ 
the residual distributions 
are used for
\begin{eq}{ll}
\hbox{nondecomposable group re\-pre\-sen\-ta\-tions}&\iff N\ge0\cr
\hbox{irreducible  group re\-pre\-sen\-ta\-tions}&\iff N=0\cr
\hbox{tangent re\-pre\-sen\-ta\-tions (below)}&\iff N\le0\cr
\end{eq}For compact groups 
strictly positive nildimensions cannot occur,
they define no functions
on the group
\begin{eq}{l}
\hbox{for compact groups }N\le 0
\end{eq}Residual distributions with 
strictly negative nildimensions $N=-1,-2,\dots$
do not lead to $G$-re\-pre\-sen\-ta\-tion matrix elements.
They
arise only for groups where the rank is strictly smaller than the 
dimension.

Associated to the dipole distribution of an irreducible 
re\-pre\-sen\-ta\-tion of the spin group  $\SU(2)$ 
and - for compact re\-pre\-sen\-ta\-tions - of the position  
manifold  $\SD(2)$ are the following distributions 
\begin{eq}{rl}
\{{1\over i\pi^2}{\rvec q~\Ga(2+N)\over (\rvec q^2\mp io-m^2)^{2+N}}\}
\hbox{ with}
&\left\{\begin{array}{ll}
m\in\Z&\hbox{for }\irrep\SU(2)\cr
m\in\R&\hbox{for }\irrep^{(2,0)}\SD(2)\end{array}\right.\cr

\int{d^3 q\over i\pi^2}{\rvec q~\Ga(2+N)\over
(\rvec q^2\mp io-m^2)^{2+N}}e^{-i\rvec x\rvec q}
&=2{\rvec x\over r }
{\p\over\p r }
({\p\over\p m^2})^{1+N}{e^{\pm i r| m|}\over r }\cr
&=-{\rvec x\over r }\left\{\begin{array}{l}
2{1\mp i r| m|\over
r^2}e^{\pm i r| m|},~ e^{\pm i r| m|},~\dots\cr
\hbox{for }N=-1,0,\dots \cr\end{array}\right.\cr

\pm \int{d^3 q\over i\pi^2}{|m|~\Ga(2+N)\over
(\rvec q^2\mp io-m^2)^{2+N}}e^{-i\rvec x\rvec q}
&=\mp 2i|m|
({\p\over\p m^2})^{1+N}
{e^{\pm i r| m|}\over r }\cr
&=\left\{\begin{array}{l}
\mp 2i|m|{e^{\pm i r| m|}\over r },~
e^{\pm i r| m|},~-{1\mp i r| m|\over 2m^2}
e^{\pm i r| m|},\dots\cr
\hbox{for }N=-1,0,1,\dots \cr\end{array}\right.\cr
\end{eq}and  to  an irreducible noncompact 
position $\SD(2)$-re\-pre\-sen\-ta\-tion
\begin{eq}{rl}
 \{{1\over i\pi^2}{\rvec q~\Ga(2+N)\over (\rvec q^2+m^2)^{2+N}}\}
 \hbox{ with}
&m\in\R
\hbox{ for }\irrep^{(1,1)}\SD(2)\cr
\int{d^3 q\over i\pi^2}{ \rvec q~\Ga(2+N)\over
(\rvec q^2+m^2)^{2+N}}e^{-i\rvec x\rvec q}&=
2{\rvec x\over r }
{\p\over\p r }\(-{\p \over \p m^2}\)^{1+N} 
{e^{- r| m|}\over  r }\cr
&=-{\rvec x\over r }\left\{\begin{array}{l}
2{1+  r| m|\over  r^2}e^{- r| m|},~ 
e^{- r| m|},\dots\cr
\hbox{for } N=-1,0,\dots\end{array}\right.\cr

\int{d^3 q\over \pi^2}{|m|~\Ga(2+N)\over
(\rvec q^2+m^2)^{2+N}}e^{-i\rvec x\rvec q}&=
2|m|\(-{\p\over \p m^2}\)^{1+N} 
{e^{- r| m|}\over  r }\cr
&=\left\{\begin{array}{l}
2|m|{e^{- r| m|}\over  r },~e^{- r| m|},~
{1+ r| m|\over 2m^2}e^{- r| m|},\dots\cr
\hbox{for }N=-1,0,1,\dots\cr\end{array}\right.\cr

\end{eq}The distributions with $N=-1$
 lead to spherical waves ${e^{\pm i r| m|}\over  r }$
and  Yukawa potentials ${e^{- r| m|}\over  r }$ and their derivatives
which are  no $\SU(2)$ and $\SD(2)$-re\-pre\-sen\-ta\-tion
matrix elements.

The distributions  associated to 
 an irreducible $\D(2)$-re\-pre\-sen\-ta\-tion
include as negative nildimension distributions
\begin{eq}{l}
{q~\Ga(3+N_0+N_3)\over (q^2_\ro P-m_0^2)^{1+N_0}
(q^2_\ro P-m_3^2)^{2+N_3}}\then\left\{\begin{array}{ccl}
{q\over (q^2_\ro P-m_0^2)(q^2_\ro P-m_3^2)},&
{q\over (q^2_\ro P-m_3^2)^2},&N_0+N_3=-1\cr
{q\over q^2_\ro P-m_0^2},&
{q\over q^2_\ro P-m_3^2},&N_0+N_3=-2\cr
\end{array}\right.
\end{eq}

\section{Residual Subrepresentations}

A re\-pre\-sen\-ta\-tion of a symmetric space (Lie group) $G$ contains 
re\-pre\-sen\-ta\-tions of subspaces (subgroups) $H$.
How does this look for residual re\-pre\-sen\-ta\-tions?

A residual $G$-re\-pre\-sen\-ta\-tion 
with 
tangent space (Lie algebra) parameters $x=(x_H,x_\perp)$ 
\begin{eq}{rl}
D^I:G&\map\C,~~g( x)\mape D^I( x)=\int d^n q~I(q)e^{iq x}
\end{eq}is projected to a
residual $H$-re\-pre\-sen\-ta\-tion 
by integration $\int d^{n-s} x_\perp$ 
over the complementary  space $\log G/H$
\begin{eq}{rl}
D^I_H:H&\map\C,~~h( x_H)\mape D_H^I( x_H)\cr
\hbox{with } D_H^I( x_H)&=
\int {d^{n-s} x_\perp\over(2\pi)^{n-s}}\int d^n q~I(q) e^{i x q}
=\int d^s q_H ~I(q_H,0) e^{i x_H q_H}
\end{eq}With the integration one picks up the Fourier components
for trivial tangent space forms (momenta) $q_\perp=0$ of $\log G/H$.

\subsection
{$\SO(2)\x \SO_0(1,1)$-Subrepresentations  in 
Spin-Position-Representations}

The $\SO(2)$-subre\-pre\-sen\-ta\-tions 
in  spin $\SU(2)$-re\-pre\-sen\-ta\-tions are given as follows 
\begin{eq}{l}
\irrep\SU(2)\map \rep\SO(2) ,~~d^2x_\perp=d^2 x_{1,2}\cr
\end{eq}
\begin{eq}{rl}
\int {d^2x_\perp\over4\pi}\int{d^3 q\over i\pi^2}{\rvec q~\Ga(2+N)\over
(\rvec q^2\mp io-m^2)^{2+N}}e^{-i\rvec x\rvec q}
&=\int{d q\over i\pi}{q~\Ga(2+N)\over
(q^2\mp io-m^2)^{2+N}}e^{-i x_3 q}\cr
&=-\ep(x_3)({\p\over\p m^2})^{1+N}
e^{\pm i| x_3 m|}\cr

\pm\int {d^2x_\perp\over4\pi}\int{d^3 q\over i\pi^2}{|m|~\Ga(2+N)\over
(\rvec q^2\mp io-m^2)^{2+N}}e^{-i\rvec x\rvec q}
&=\pm \int{d q\over i\pi}{|m|~\Ga(2+N)\over
(q^2\mp io-m^2)^{2+N}}e^{-i x_3 q}\cr
&=|m|({\p\over\p m^2})^{1+N}
{e^{\pm i| x_3 m|}\over|m|}\cr
\end{eq}and the $\SO_0(1,1)$-sub\-re\-pre\-sen\-ta\-tions
of noncompact position $\SD(2)$-re\-pre\-sen\-ta\-tions 
\begin{eq}{l}
\irrep\SD(2)\map\rep^{(1,1)}\SO_0(1,1)\cr
\end{eq}
\begin{eq}{rl}
\int {d^2x_\perp\over4\pi}\int{d^3 q\over i\pi^2}{\rvec q~\Ga(2+N)\over
(\rvec q^2+m^2)^{2+N}}e^{-i\rvec x\rvec q}
&=\int{d q\over i\pi}{q~\Ga(2+N)\over
(q^2+m^2)^{2+N}}e^{-i x_3 q}\cr
&=-\ep(x_3)(-{\p\over\p m^2})^{1+N}e^{-| x_3 m|}\cr

\int {d^2x_\perp\over4\pi}\int{d^3 q\over \pi^2}{|m|~\Ga(2+N)\over
(\rvec q^2+m^2)^{2+N}}e^{-i\rvec x\rvec q}
&=\int{d q\over \pi}{|m|~\Ga(2+N)\over
(q^2+m^2)^{2+N}}e^{-i x_3 q}\cr
&=|m|(-{\p\over\p m^2})^{1+N}{e^{-| x_3 m|}\over|m|}\cr
\end{eq}The vector dependence ${\rvec x\over  r }$
for the sphere is projected to two values $\ep(x_3)\in\{\pm1\}$
for the hemispheres.

\subsection
{Time and Position Subrepresentations in Spacetime Representations}

The energy-momentum  distribution used 
in the residual spacetime re\-pre\-sen\-ta\-tions is
the principal value part in the decomposition of a complex 
distribution into  imaginary and real part
\begin{eq}{l}
\pm{1\over i\pi}
{q \over   (q^2\mp io-m_0^2)(q_\ro P^2-m_3^2)^2}
={\scriptsize
\pm\underbrace{ {1\over i \pi}
{q\over   (q_\ro P^2-m_0^2)(q_\ro P^2-m_3^2)^2}}
_{{\rm spacetime}~\D(2)~\to~\C}
+\underbrace{ {1\over(m_0^2-m_3^2)^2}
q\de(q^2-m_0^2) }_{{\rm tangent ~translations}~ \R^4\to~\C}
} 
\end{eq}which is also the decomposition for the 
re\-pre\-sen\-ta\-tion matrix elements of 
 spacetime $\D(2)$ and its tangent space $\R^4$.
The integrated principal value  part
 has causal support whereas
the integrated  Dirac distribution  for the particle pole  
gets both spacelike and causal  support.
The decomposition with respect to the two singularities 
\begin{eq}{rl}
{1\over(q ^2- m_0^2)(q^2- m_3^2)^2}&=
{1\over (m_0^2-m_3^2)^2}\Brack{ {1\over q^2-m_0^2}
-
{q^2-m_0^2\over (q^2- m_3^2)^2} } \cr
&={1\over (m_0^2-m_3^2)^2}\( {1\over q^2- m_0^2}
-{1\over q^2- m_3^2}\)
-{1\over m_0^2-m_3^2}{1\over (q^2- m_3^2)^2}\cr
\end{eq}is not parallel with 
the re\-pre\-sen\-ta\-tion of the factors in
$\D(2)=\D(\bl1_2)\x\SD(2)$.
The projections to  re\-pre\-sen\-ta\-tion matrix elements  of
the manifold factors  are given 
by  { position  integration for the causal group
 $\D(\bl1_2)$} 
 and by
{  time integration for the position   manifold} $\SD(2)$
with Cartan subgroup $\SO_0(1,1)$,
 i.e. by the Fourier transforms for
trivial momenta $\rvec q=0$ and trivial energy $q_0=0$ resp. 
\begin{eq}{rlrl}
\int d^3 x :&\irrep\D(2)&\map&\rep \D(1)\cr 
\int d x_0:&\irrep\D(2)&\map&\rep \SD(2)\cr 
&&\int d^2 x_\perp:&\rep\SD(2)\map\rep \SO_0(1,1)\cr 
\end{eq}where one uses
\begin{eq}{rlrl}
\left({\scriptsize\begin{array}{r}
\int{d^3x\over 8\pi}\cr
\int{dx_0\over 2}\cr
\int{d^2x_\perp\over4\pi}\int{dx_0\over 2}\cr\end{array}}\right)
\int {d^4q\over i\pi^3}{q~\Ga(3+N)\over (q^2_\ro P-m^2)^{3+N} }e^{xiq}
&=&({\p\over \p m^2})^{2+N}
&{\scriptsize\pmatrix{
\ep(x_0)\cos x_0m\cr
2{\rvec x\over r }
{1+ r|m|\over r^2}e^{- r|m|}\cr
-\ep(x_3)e^{-|x_3m|}\cr}}\cr
\left({\scriptsize\begin{array}{r}
\int{d^3x\over 8\pi}\cr
\int{dx_0\over 2}\cr
\int{d^2x_\perp\over4\pi}\int{dx_0\over 2}\cr\end{array}}\right)
\int {d^4q\over \pi^3}{\Ga(2+N)\over (q^2_\ro P-m^2)^{2+N} }e^{xiq}
&=&-({\p\over \p m^2})^{1+N}&
{\scriptsize\pmatrix{
{\sin |x_0m|\over|m|}\cr
2{e^{- r|m|}\over r }\cr
{e^{-|x_3m|}\over|m|}\cr}}\cr
\end{eq}This leads for  irreducible spacetime  
re\-pre\-sen\-ta\-tions to
\begin{eq}{rl}\int {d^3 x\over 8\pi} (m_0^2;1,-m_3^2)(x)
&=\int {dq_0\over i\pi}
{q_0 \over   (q_{0\ro P}^2 -m_0^2)(q_{0\ro P}^2-m_3^2)^2}
    e ^{x_0iq_0}\cr
&=\ep(x_0)\Brack{
{\cos x_0 m_0-\cos x_0 m_3\over(m_0^2-m_3^2)^2}
+{|x_0|\sin |x_0m_3|\over 2|m_3|(m_0^2-m_3^2)} }\cr
\int {d x_0\over2} (m_0^2;1,-m_3^2)(x)
&= \int
{d^3q\over i\pi^2} {\rvec q \over   (\rvec q^2+m_0^2)
(\rvec q^2+m_3^2)^2}     e ^{-\rvec xi\rvec q}\cr
&=-{\rvec x\over r }   \Brack{
2
{(1+ r| m_0|)  e ^{- r| m_0|}
-(1+ r| m_3|)  e ^{- r| m_3|}
\over r^2(m_0^2-m_3^2)^2}
+{e ^{- r| m_3| }\over m_0^2-m_3^2}
  }\cr
\int {d^2x_\perp\over 4\pi}\int {d x_0\over2}(m_0^2;1,-m_3^2)(x)
&= \int
{dq\over i\pi} {q \over   (q^2+m_0^2)
(q^2+m_3^2)^2}     e ^{-x_3iq}\cr
&=-\ep(x_3)\Brack{
{e ^{-|x_3 m_0|}- e ^{-|x_3 m_3|}
\over (m_0^2-m_3^2)^2}
+{|x_3|e ^{- r| m_3| }\over 2|m_3|(m_0^2-m_3^2)}
}
 \end{eq}

The measure of the invariants  for 
 an irreducible spacetime re\-pre\-sen\-ta\-tion
\begin{eq}{l}
\hbox{for }\D(2):~~\rho(M_0^2,M_3^2)=\de(M_0^2-m_0^2)\de(M_3^2-m_3^2)
\end{eq}is projected to measures for the re\-pre\-sen\-ta\-tion of the two factors.
The time $\bl D(\bl 1_2)$-subre\-pre\-sen\-ta\-tion with the measure
\begin{eq}{l}
\hbox{for }\D(\bl1_2):~\rho_0(m^2)=
{\de(m^2- m_0^2)
-\de(m^2- m_3^2)\over (m_0^2-m_3^2)^2}
+{\de'(m^2- m_3^2)\over m_0^2-m_3^2}
\end{eq}contains matrix elements of reducible nondecomposable re\-pre\-sen\-ta\-tions
for the nonparticle dipole at $m_3^2$.

The linear combinations occurring
in the position   $\SD(2)$-projections 
of spacetime $\D(2)$-re\-pre\-sen\-ta\-tions
are matrix elements of measured $\SD(2)$-re\-pre\-sen\-ta\-tions
involving the difference  of two Yukawa potentials 
\begin{eq}{rl}
2
{e^{- r| m_0|}-e^{- r| m_3|}\over r }
=&\int _{m_0^2}^{m_3^2}dm^2
{e^{- r| m|}\over |m|}=\int_0^\infty dm^2
\vth(m^2-m_0^2)\vth(m_3^2-m^2)
{e^{- r| m|}\over |m|}\cr
\end{eq}The measure for the $\SD(2)$-sub\-re\-pre\-sen\-ta\-tion reads
\begin{eq}{l}
\hbox{for }\SD(2):~\rho_3(m^2)
=-{
\vth(m^2- m_0^2)\vth(m_3^2- m^2)\over (m_0^2-m_3^2)^2}
+{\de(m^2- m_3^2)\over m_0^2-m_3^2}
\end{eq}

\section
{Residual Tangent Distributions}

The {residual tangent distributions}
for an irreducible symmetric space (group) re\-pre\-sen\-ta\-tion
will be defined by the {associated distributions
with a simple pole}, i.e. for minimal negative nildimension $N\le0$,
and a trivial invariant.
They arise as the inverse differential operators in the Lie algebra action
representing differential equations of motions.

The tangent $\R$ distributions
for the Abelian groups have trivial nildimensions $N=0$ - for the non-selfdual
ones
\begin{eq}{l}
\left.\begin{array}{l}
\log\U(1):\cr\log\D(1):\cr\end{array}\right\}
~~
\de(q)\cong{1\over2i\pi}{1\over q},~~
\oint{dq\over2i\pi}{1\over q}e^{tiq}=1
\end{eq}for the selfdual compact representations
\begin{eq}{l}
\left.\begin{array}{l}
\log\SO(2):\cr\log\SO_0(1,1):\cr\end{array}\right\}~~
{1\over i\pi}{q\over
q^2\mp io},~~
\int	{d^1q\over i\pi}{q\over
q^2\mp io}e^{tiq}=\ep(t)\cr
\end{eq}and for the selfdual noncompact $\SO_0(1,1)$-representations
\begin{eq}{l}
\log\SO_0(1,1):~~
{1\over i\pi}{q\over
q^2+o^2},~~
\int	{d^1q\over i\pi}{q\over
q^2+o^2}e^{-xiq}=-\ep(x)\cr\end{eq}

For the nonabelian rank 1 spaces the residual tangent $\R^3$ distributions
come with nildimension $N=-1$ 
\begin{eq}{l}
\left.\begin{array}{l}
\log\SU(2):\cr\log\SD(2):\cr\end{array}\right\}
~~
{1\over i\pi^2}{\rvec q\over
\rvec q^2\mp io-m^2},~~
\int {d^3q\over i\pi^2}{\rvec q\over
\rvec q^2\mp io}e^{-\rvec xi\rvec q}=-2{\rvec x\over r^3}
\end{eq}and in the noncompact 
case
\begin{eq}{l}
\log\SD(2):
~~
{1\over i\pi^2}{\rvec q\over
\rvec q^2+o^2},~~
\int {d^3q\over i\pi^2}{\rvec q\over
\rvec q^2+o^2}e^{-\rvec xi\rvec q}=-2{\rvec x\over r^3}
\end{eq}They lead both to the Coulomb force with the Cartan subalgebra
projection
\begin{eq}{l}
\int{d^2x_\perp\over 4\pi}2{\rvec x\over r^3}=\ep(x_3)
\end{eq}

The residual tangent spacetime $\R^4$ distributions
have nildimension $N_0+N_3=-2$ 
\begin{eq}{l}
\log\D(2):~~
{1\over i\pi^3}{q\over q^2_\ro P},~~
\int {d^4q\over i\pi^3}{q\over q^2_\ro P}e^{xiq}
={x\over2}\de'({x^2\over4})\cr
\hbox{with projections }
{\scriptsize\pmatrix{
\int{d^3 x\over 8\pi}\cr
\int{d x_0\over 2}\cr
\int{d^2x_\perp\over 4\pi}\int{d x_0\over 2}\cr}}
\int {d^4q\over i\pi^3}{q\over q^2_\ro P}e^{xiq}
={\scriptsize\pmatrix{
\ep(x_0)\cr
2{\rvec x\over r^3}\cr
\ep(x_3)\cr}}
\end{eq}

\section{Defining Representations for Time, Position and Spacetime}

Spacetime, particles and interactions cannot be taken as
separate concepts. Spacetime is known via interacting particles
and the interactions
of  particles can be  understood only in spacetime.

This connection will be translated into the mathematical language with the 
concept of a defining representation,
familiar from Lie groups. E.g., the Lie group $\SU(n)$ is defined by
the automorphisms of a vector space $V\cong\C^n$ 
compatible with a scalar product  -
the linear space and the operating group merge in the concept 
of the defining representation.

In addition to one defining representation for some  Lie groups there exist
fundamental representations which reflect the rank
and the number of independent invariants. E.g.,
the Lie symmetry $\SU(r+1)$ one has $r$ fundamental representations
whose highest weights are basic vectors for the $\Z$-module with all weights. 
The products of  a defining representation may build the fundamental ones,
e.g. in the case of $\SU(n)$ via 
the totally antisymmetric Grassmann powers of the
defining vector space. 

\subsection{The Harmonic Oscillator - Defining a Compact Time}

The irreducible  time $\D(1)$ representation in the group
 $\U(1)$ as seen in
the quantization
 for creation and annihilation
operators $(\ro u,\ro u^*)$ of a harmonic Fermi or Bose oscillator
with frequencies $\pm m\in\R$ 
\begin{eq}{l}
\D(1)\ni e^t\mape 
e^{tim}
=[\ro u^*,\ro u]_\pm
\in\U(1)
\end{eq}defines a compact model for time 
with the invariant 
${1\over |m|}$ 
as characteristic time unit.

The adjoint action with the Hamiltonian
as the represented Lie algebra   basis 
defines the time translations in the equations of motion
\footnote{\scriptsize
 $\ro u$ without argument
means $\ro u(0)$, i.e. for the trivial translation.}
\begin{eq}{l}
H=
m{\com{\ro u}{\ro u^*}_\mp\over 2}
\then\left\{\begin{array}{llll} 
{d\ro u\over dt}&=[iH,\ro u]&=im\ro u,&\ro u(t)=e^{tim}\ro u\cr
{d\ro u^*\over dt}&=[iH,\ro u^*]&=-im\ro u^*,&\ro u^*(t)=e^{-tim}\ro u^*\cr
\end{array}\right.
\end{eq}The operators are $\U(1)$-isomorphic time orbits in the
$\C$-isomorphic representation spaces
\begin{eq}{l}
\ro u,\ro u^\star:\D(1)\map V,V^T\cong\C
\end{eq}

The product representations
$ e^{tim_1}e^{tim_2}=e^{ti(m_1+m_2)}$
generate the familiar equidistant 
time weights (eigenvalues, frequencies)
for the quantum oscillator - $\{Zm\mid Z\in\Z\}$ for Bose 
and $\{Zm\mid Z=0,\pm1\}$ for Fermi which - for the states - are projected on
the positive values.

\subsection{The Exponential Potential - Defining a Noncompact Position}

An indefinite unitary representation of the noncompact Procrustes dilatation
group $\SO_0(1,1)$ for dual 
operators $(\ro d,\ro d^*)$  of  Fermi or Bose type
with eigenvalues $\pm m\in\R$
\begin{eq}{l}
\SO_0(1,1)\ni  
{\scriptsize\pmatrix{e^{-x}&0\cr 0&e^{x}\cr}}\mape
{\scriptsize\pmatrix{e^{-xm}&0\cr 0&e^{xm}\cr}}
={\scriptsize\pmatrix{
[\ro d^*,\ro d]_\pm &[\ro d,\ro d]_\pm\cr
[\ro d^*,\ro d^*]_\pm&\pm[\ro d,\ro d^*]_\pm\cr}}(x)\in\SU(1,1)
\end{eq}defines a faithful model for the position space Cartan
subgroup $\SO_0(1,1)$ with the invariant 
${1\over |m|}$ as characteristic length unit.

The translations are 
implemented with the basis
\begin{eq}{l}
D=
im{[\ro d,\ro d^*]_\mp\over2}
\then\left\{\begin{array}{llll}{d\ro d\over dx}&=[iD,\ro d]&=-m\ro d,&
\ro d(x)=e^{-xm}\ro d\cr
{d\ro d^*\over dx}&=[iD,\ro d^*]&=m\ro d^*,&\ro d^*(x)=e^{xm}\ro d^*\cr
\end{array}\right.
\end{eq}The operators are noncompact $\D(1)$-isomorphic dilatation orbits
in the
$\C$-isomorphic representations spaces
\begin{eq}{l}
\ro d,\ro d^*:\SO_0(1,1)\map V,V^T\cong\C
\end{eq}

The product representations (convolutions)
lead to exponentials with the eigenvalues
 $\{zm\mid z=0,\pm1\}$ for Fermi 
and  $\{zm\mid z\in\Z\}$ for Bose.

A  representation matrix element of 
the symmetric space position model 
$\SD(2)\cong\SL(\C^2)/\SU(2)$
\begin{eq}{l}
\SD(2)\ni 
e ^{-\rvec x \rvec \si}
\mape -{\rvec\si\rvec x\over r }  e ^{-r |m|}
=\int {d^3 q\over i\pi^2}{\rvec\si\rvec q\over (\rvec q^2+m^2)^2}
e ^{-\rvec x i\rvec q} =\acom{\psi^*}\psi(\rvec x)
\end{eq}with Pauli matrices $\rvec \si$
defines  a noncompact position with a characteristic length
${1\over |m|}$
(interaction range), implemented by 
$\C^2$-valued Pauli spinor fields 
on the position manifold
\begin{eq}{l}
\psi^A,\psi^*_A:\SD(2)\map V,V^T\cong\C^2,~~A=1,2
\end{eq}The Cartan subgroup $\SO_0(1,1)$ is represented
by  an indefinite unitary $\SU(1,1)$-representation matrix element
$e ^{-r|m|}$.

The product representations (convolutions)
add up the noncompact invariants $\{n|m|\mid n=1,2,\dots\}$ 
in the exponential and
are multiplied  with  spherical harmonics 
of degree $\{2J\mid 2J=0,1,2,\dots\}$ for the representation of the
sphere $\SO(3)/\SO(2)$.

\subsection{Defining Spacetime with Two Invariants}

The representation matrix element
\begin{eq}{l}
\D(2)\ni \vth(x^2) x\mape \int {d^4 q\over i\pi^3}
{2\si^jq_j \over   (q_\ro P^2-m_0^2)(q_\ro P^2-m_3^2)^2}e^{xiq}
=\ep(x_0)\acom{\bl \Psi^*}{\bl \Psi}(x)
\end{eq}defines  symmetric spacetime\cite{S97}. The two invariants
$m_0^2$ and $m_3^2$ characterize time and position 
and give units for particle masses
and interaction lengths. The representation is implemented by 
$\C^2$-valued Weyl spinor fields\cite{HEI}
\begin{eq}{l}
\bl\Psi^A,\bl\Psi^*_{\dot A}:\D(2)\map V,V^T\cong\C^2,~~A=1,2
\end{eq}

It involves two conjugations - a definite 
$\U(2)$-conjugation for the time $\D(1)$-representation
and an indefinite 
$\U(1,1)$-conjugation for the position $\SD(2)$-representation.
Therefore only the particle pole can be endowed with an additional
asymptotic positive unitary spacetime translation representation 
structure by adding a real
on shell contribution via $\pm {1\over i\pi}{1\over   q^2\mp io-m_0^2}$.
 A parametrization with creation and annihilation operators
has to take care of the indefinite conjugation involved.

The product representations of the defining spacetime
representation  will give rise 
to product invariants which - in the case
of an accompanying definite unitary conjugation, can be identified with
particle masses for
 bound states. To carry out explicitly such a program, i.e.
to compute a mass spectrum from the spacetime defining two invariants, the
representation characteristic ratio ${m_0^2\over m_3^2}$ 
has to be determined as
well as the relevant normalization factors to be used in the
eigenvalue equations for the product representation invariants.


\end{document}